\documentclass[a4paper,11pt]{article}
\usepackage{jcappub}

\usepackage{booktabs}
\usepackage[english]{babel}
\usepackage{amsmath,amssymb,amsbsy,amstext, amsthm, simplewick}
\usepackage{hyperref}
\usepackage{graphicx}
\usepackage{amsfonts}
\usepackage{amssymb}
\usepackage{framed}
\usepackage{enumitem}
\usepackage{soul}
\usepackage{orcidlink}
\usepackage{upgreek}
 \usepackage{exscale,relsize}
 \usepackage[makeroom]{cancel}
\usepackage{soul}
\usepackage{bbold}
\usepackage{lipsum}
\usepackage{mdframed}
\usepackage{mathtools}
\usepackage[dvipsnames]{xcolor}
\usepackage{tikz}
\usepackage{tikz-cd}
\usepackage[export]{adjustbox}
\usepackage{dsfont}
\usepackage[mathscr]{eucal}
\usepackage{comment}

 \newcommand{\circled}[2][]{%
  \tikz[baseline=(char.base)]{%
    \node[shape = circle, draw, inner sep = 1pt]
    (char) {\phantom{\ifblank{#1}{#2}{#1}}};%
    \node at (char.center) {\makebox[0pt][c]{#2}};}}
\robustify{\circled}

\usepackage{colortbl}
\definecolor{lightgreen}{cmyk}{0.2, 0, 0.2, 0.2}
\definecolor{lightgray}{cmyk}{0.1,0.2,0,0.1}
\definecolor{lightgray2}{cmyk}{0.1,0.1,0,0.1}

\makeatletter
\newlength{\apb@width}
\newcommand{\autoparbox}[2][c]{\settowidth{\apb@width}{#2}\parbox[#1]{\apb@width}{#2}}

\newcommand{\Cen}[2]{%
  \ifmeasuring@
    #2%
  \else
    \makebox[\ifcase\expandafter #1\maxcolumn@widths\fi]{$\displaystyle#2$}%
  \fi
}
\makeatother

\newcommand{\beq}{\begin{equation}\begin{aligned}}
\newcommand{\eeq}{\end{aligned}\end{equation}}

\RequirePackage{amsmath}
\RequirePackage{amssymb}
\RequirePackage{epsfig}
\RequirePackage{graphicx}
\RequirePackage[numbers,sort&compress]{natbib}
\RequirePackage[colorlinks=true
  ,urlcolor=blue
  ,anchorcolor=blue
  ,citecolor=blue
  ,filecolor=blue
  ,linkcolor=blue
  ,menucolor=blue
  ,pagecolor=blue
  ,linktocpage=true
  ,pdfproducer=medialab
  ,pdfa=true
]{hyperref}

\def\beq{\begin{equation}}
\def\eeq{\end{equation}}

\def\Beq{\begin{equation}\begin{aligned}}
\def\Eeq{\end{aligned}\end{equation}}

\def\bea{\begin{eqnarray}}
\def\eea{\end{eqnarray}}

\def\beq{\begin{equation}}
\def\eeq{\end{equation}}
\def\bea{\begin{eqnarray}}
\def\eea{\end{eqnarray}}

\def\bk{\boldsymbol{k}}

\DeclareRobustCommand{\SkipTocEntry}[4]{}
\DeclareSymbolFont{extraup}{U}{zavm}{m}{n}
\DeclareMathSymbol{\varheart}{\mathalpha}{extraup}{86}
\DeclareMathSymbol{\vardiamond}{\mathalpha}{extraup}{87}
\title{Gravitational Waves from Matter Perturbations of Spectator Scalar Fields}
\author[a, 1]{Marcos A. G. Garcia \orcidlink{0000-0003-3496-3027}}
\affiliation[a  ]{Departamento de F\'isica Te\'orica, Instituto de F\'isica, \\
Universidad Nacional Aut\'onoma de M\'exico, \\
Ciudad de M\'exico C.P. 04510, Mexico} 

\author[a, 2]{\'Angel Garc\'ia-Vega \orcidlink{0009-0000-9754-7739}}

\author[b, 3]{Sarunas Verner \orcidlink{0000-0003-4870-0826}}
\affiliation[b]{Kavli Institute for Cosmological Physics, \\
University of Chicago, 5640 South Ellis Ave., Chicago, IL 60637, USA}

\emailAdd{marcos.garcia@fisica.unam.mx}
\emailAdd{angelgarcia@estudiantes.fisica.unam.mx}
\emailAdd{verner@uchicago.edu}

\abstract{We compute the stochastic gravitational wave background sourced at second order by a spectator scalar field $\chi$ coupled to the inflaton $\phi$ through a portal interaction $\sigma\phi^2\chi^2$ and with quartic self-interaction $\lambda_\chi\chi^4/4!$. In the large portal coupling regime ($\sigma/\lambda \gg 1$, with $\lambda$ the inflaton normalization), parametric resonance during reheating amplifies the spectator power spectrum by many orders of magnitude near the resonance band until Hartree backreaction from the quartic coupling detunes the instability, while the large inflationary effective mass suppresses superhorizon power and ensures compatibility with CMB isocurvature bounds. We focus on the direct field-gradient source $\partial_a\chi\,\partial_b\chi$ in the second-order Einstein equations and derive a master formula that factorizes into a spectral integral over the frozen, vacuum-subtracted spectator spectrum and a time integral encoding the post-inflationary expansion history. For our benchmark reheating history we obtain analytic scaling relations, including a peak amplitude $\Omega_{\rm GW}\propto T_{\rm reh}^{8/3}$, strong dependence on the portal strength, and weak sensitivity to $m_\chi$. We validate the framework against nonlinear lattice simulations, demonstrating complementarity: the Hartree treatment captures superhorizon evolution inaccessible to the lattice, while the lattice resolves rescattering and fragmentation near the spectral peak. For $\sigma/\lambda \simeq 10^4$ and $T_{\rm reh}=2 \times 10^{14}\,\mathrm{GeV}$, the signal reaches $\Omega_{\rm GW}h^2\sim 10^{-11}$ at $f\sim10^{7}$--$10^{8}\,\mathrm{Hz}$. Increasing $\lambda_\chi$ at fixed $\sigma$ has a non-monotonic effect: small values enhance the signal via rescattering, whereas larger values suppress it by detuning the resonance.}

\begin{document}
\maketitle
\flushbottom

%%%%%%%%%%%%%%%%%%%%%%%%%%%%%%%%%%%%%%%%%%%%%%%%
\section{Introduction}
\label{sec:introduction}
%%%%%%%%%%%%%%%%%%%%%%%%%%%%%%%%%%%%%%%%%%%%%%%%
Gravitational waves (GWs) offer a unique window into the earliest epochs of the universe, probing energy scales and dynamics that are inaccessible to electromagnetic observations~\cite{Caprini:2018mtu,Maggiore:1999vm}. Recent pulsar timing array results~\cite{NANOGrav:2023hvm,EPTA:2023fyk, Reardon:2023gzh, Xu:2023wog} have reported evidence consistent with a stochastic GW background, and upcoming CMB polarization experiments such as LiteBIRD~\cite{Hazumi:2019lys} and the Simons Observatory~\cite{SimonsObservatory:2018koc} will soon test inflationary models through primordial $B$-mode searches. Future missions like PICO~\cite{NASAPICO:2019thw} are also being proposed to improve sensitivity. These advances heighten the importance of identifying complementary GW signatures of inflation and reheating beyond the standard tensor-to-scalar ratio.

A promising and well-motivated source of such signatures is the class of \emph{spectator} scalar fields: fields that are energetically subdominant during inflation but acquire primordial fluctuations from the quasi-de Sitter background. Depending on the ultraviolet completion, a spectator may correspond to a stable dark matter candidate~\cite{Chung:1998zb,Chung:1998rq,Peebles:1999fz,Byrnes:2006fr,Markkanen:2018gcw,Kolb:2023ydq,Ling:2021zlj,Garcia:2022vwm,Garcia:2025rut}, the Standard Model Higgs~\cite{Herranen:2014cua,Espinosa:2007qp},
a curvaton that decays at late times~\cite{Linde:1996gt,Enqvist:2001zp,Moroi:2001ct,Lyth:2001nq}, or a Peccei-Quinn scalar associated with the strong CP problem~\cite{Peccei:1977ur,Peccei:1977hh,Bao:2022hsg,Chen:2023txq}. On CMB scales, \textit{Planck} limits require the uncorrelated CDM isocurvature power at $k_*=0.05~{\rm Mpc}^{-1}$ to be suppressed to $\Delta_{\mathcal S}^2(k_*)\lesssim{\cal O}(10^{-10})$~\cite{Planck:2018jri}. However, these constraints do not preclude a spectrum that grows rapidly at shorter wavelengths. In particular, a blue-tilted spectator spectrum can become substantial on scales $k\gg k_*$, beyond the reach of present CMB measurements, with potentially observable secondary signatures~\cite{Domenech:2021ztg, Ebadi:2023xhq,Garcia:2023awt,Garcia:2025yit,Garcia:2025wmu}.

In our previous work~\cite{Ebadi:2023xhq,Garcia:2023awt,Garcia:2025yit,Garcia:2025wmu},
we showed that blue-tilted spectator spectra arise naturally when the effective mass satisfies
$m_{\chi,\rm eff}\gtrsim 0.5\,H_*$~\cite{Chung:2004nh}, and that the resulting second-order GW background can strengthen this bound to $m_{\chi,\rm eff}\gtrsim 0.7\,H_*$. Those studies focused on spectators with small or vanishing portal couplings, for which gravitational particle production is the dominant production
mechanism. The present work goes beyond this regime by considering large portal couplings, where parametric amplification during reheating qualitatively changes the dynamics and produces a far stronger GW signal. More broadly, cosmological perturbations on small scales ($k\gg 1~{\rm Mpc}^{-1}$) provide a largely untapped probe of physics between the end of inflation and Big Bang nucleosynthesis. Even a modest enhancement of the primordial spectrum can produce observable consequences such as ultracompact minihalos testable with gravitational lensing~\cite{DES:2020fxi,Boddy:2022knd}, and if the scalar perturbations reach $\Delta_\zeta^2\gtrsim 10^{-7}$ over some band of scales, a detectable stochastic GW background is generated at second order~\cite{Kohri:2018awv,Domenech:2021ztg}. The spectator mechanism studied here provides a natural and well-motivated route to such enhancements.

In this work, we consider a spectator $\chi$ described by the action
\begin{equation}
\label{eq:generalaction}
    \mathcal{S}_\chi
    \;=\;
    \int d^4x\sqrt{-g}\left[
    -\frac{1}{2}(\partial_\mu\chi)^2
    -\frac{1}{2}m_\chi^2\chi^2
    -\frac{1}{2}\sigma\phi^2\chi^2
    -\frac{\lambda_\chi}{4!}\chi^4\right],
\end{equation}
where $m_\chi$ is the bare mass, $\sigma$ controls the inflaton-spectator portal coupling, and $\lambda_\chi$ the quartic self-interaction. We focus on the strong portal regime, conveniently parameterized by $\sigma/\lambda\gg 1$ where $\lambda$ is the normalization of the inflaton potential, so that the post-inflationary dynamics differ qualitatively from the purely gravitational production operating at $\sigma\to 0$. In this regime, the oscillating inflaton condensate drives broad parametric resonance in the spectator mode equation, amplifying the spectator fluctuations by many orders of magnitude near the resonance band over a few $e$-folds of reheating~\cite{Kofman:1994rk,Shtanov:1994ce,Kofman:1997yn}. This process is closely analogous to preheating in the inflaton sector~\cite{Traschen:1990sw,Khlebnikov:1996mc,Greene:1997fu} and is also known to source stochastic GW backgrounds in a variety of models (see e.g.\ Refs.~\cite{Khlebnikov:1997di,Easther:2006gt,Easther:2006vd,Dufaux:2007pt,Garcia-Bellido:2007nns,Figueroa:2017vfa,Adshead:2018doq,Adshead:2019lbr}). Here, however, resonance populates a spectator sector that remains subdominant in the \emph{background} energy budget. The quartic self-interaction regulates the amplification through Hartree backreaction: the growing variance $\langle\chi^2\rangle$ generates an effective mass that detunes the resonance, with $\sigma$ and $\lambda_\chi$ jointly controlling the efficiency. Simultaneously, a large inflationary effective mass $\sigma\phi_*^2\gg H_*^2$ places the spectator in the white-noise regime, yielding a maximally blue frozen spectrum ($\propto k^3$ at the end of inflation) that comfortably satisfies the CMB isocurvature bound while concentrating spectral weight at small scales. For comparison, we have verified that in the purely gravitational limit ($\sigma=0$) the resulting GW signal is too small to be detectable across the reheating temperatures and parameter space consistent with our dark sector constraints~\cite{Garcia:2022vwm,Garcia:2023qab,Garcia:2025rut}, establishing parametric amplification as the essential ingredient in the regime of interest.

The parametrically amplified spectator fluctuations source GWs through the transverse-traceless (TT) projection of the second-order Einstein equations. Two distinct channels contribute. The first is the \emph{metric-sourced} ($\Phi\Phi$) channel, in which density perturbations source the Newtonian potential $\Phi$ and the quadratic $\Phi\Phi$ terms then enter the tensor equation. This channel was analyzed in Refs.~\cite{Ebadi:2023xhq,Garcia:2025yit,Garcia:2025wmu} and probes relatively large scales. The second is the \emph{direct field gradient} ($\partial\chi\,\partial\chi$) channel, arising from the spectator anisotropic stress at second order. Since the spectator does not source $\Phi$ at linear order (being subdominant with $\bar\chi=0$), this direct contribution is independent of the inflaton-sourced metric channel and is additional to it. In this paper, we present the first semi-analytical computation of the direct field-gradient contribution for a parametrically amplified spectator, providing analytical control over the parameter dependence that complements existing lattice studies of GW production from (resonant) reheating and preheating~\cite{Khlebnikov:1997di,Easther:2006gt,Easther:2006vd,Dufaux:2007pt,Garcia-Bellido:2007nns,Figueroa:2017vfa,Cosme:2022htl,Cui:2023fbg,Adshead:2018doq,Adshead:2019lbr, Garcia:2021iag, Garcia:2022vwm}.

Our central result is a master formula for the GW power spectrum that factorizes into two pieces: a spectral integral $g(k)$ over the frozen, vacuum-subtracted spectator spectrum and a time-dependent factor $\mathcal{I}(k,N)$ encoding the build-up of the GW signal through the post-inflationary expansion history. We show that a leading adiabatic (vacuum) subtraction, $|X_k|^2-1/(2\omega_k)$, is numerically essential
for ultraviolet convergence of the momentum integral. The time-dependent factor is decomposed into contributions from three distinct epochs: the parametric amplification phase, the matter-dominated reheating era, and the subsequent radiation era. For our benchmark reheating history this decomposition yields closed-form scaling relations, including $\Omega_{\rm GW}\propto T_{\rm reh}^{8/3}$ at the spectral peak, providing physical insight into which epoch dominates and how the signal depends on $T_{\rm reh}$, the portal strength (conveniently expressed through $\sigma/\lambda$), the spectator mass $m_\chi$, and the self-interaction $\lambda_\chi$.

We validate the semi-analytical framework against full nonlinear lattice simulations performed with \textsc{CosmoLattice}~\cite{Figueroa:2020rrl,Figueroa:2021yhd}, finding excellent agreement (within numerical uncertainties and over the wavenumber range accessible to the lattice) for $k\lesssim k_{\rm end}$ in
both the spectator field power spectrum and the GW spectrum. The two approaches are complementary: the Hartree calculation extends the prediction to deeply superhorizon scales that would be prohibitively expensive to simulate on the lattice, while the lattice captures mode-mode rescattering, inflaton fragmentation, and the turbulent cascade to UV modes~\cite{Micha:2002ey,Micha:2004bv,Figueroa:2023oxc} that lie beyond the mean-field approximation. For a curvaton-like spectator with $\sigma/\lambda=10^4$ and $T_{\rm reh}= 2 \times 10^{14}$~GeV, the predicted signal reaches peak amplitudes of $\Omega_{\rm GW}h^2\sim
10^{-11}$ at frequencies $f\sim 10^7 - 10^8$~Hz. The steep infrared scaling ($\Omega_{\rm GW}\propto f^5$) places the signal in the ultra high-frequency regime, beyond the sensitivity bands of current and
planned interferometric GW detectors but motivating proposed resonant-cavity and other novel experiments~\cite{Aggarwal:2020olq,Herman:2022fau,Ringwald:2022xif}. Increasing $\lambda_\chi$ at fixed $\sigma$ produces a non-monotonic effect: small values can enhance the signal through rescattering, whereas
larger values suppress it by detuning the resonance.

The remainder of this paper is organized as follows. Section~\ref{sec:infandmodel} describes the inflationary model, the spectator field action, and the post-inflationary reheating dynamics. Section~\ref{sec:gpp} discusses spectator particle production through parametric amplification, including the Bogoliubov formalism and the Hartree approximation for the self-consistent evolution of the effective mass. Section~\ref{sec:matterPS} defines the vacuum-subtracted spectator field power spectrum, discusses its evolution through reheating, and validates the Hartree calculation against lattice simulations. Section~\ref{sec:SIGWs} develops the GW formalism: the second-order tensor equation, the TT projection, the Green's function solution, the four-point function decomposition, and the master formula
for the GW power spectrum. In Section~\ref{sec:gwmat} we evaluate the time-dependent factor semi-analytically, decomposing it into contributions from the parametric amplification, reheating, and radiation eras. Section~\ref{sec:gwsignals} presents the predicted GW spectra for curvaton-like scenarios, exploring the dependence on $T_{\rm reh}$, $\sigma/\lambda$, $m_\chi$, and $\lambda_\chi$, and comparing with lattice results. We conclude in Section~\ref{sec:conclusions}. Throughout, we use natural units ($\hbar=c=k_B=1$) and the metric signature $(-,+,+,+)$.

%%%%%%%%%%%%%%%%%%%%%%%%%%%%%%%%%%%%%%%%%%
\section{Inflation and Spectator Scalar Fields}
\label{sec:infandmodel}
%%%%%%%%%%%%%%%%%%%%%%%%%%%%%%%%%%%%%%%%%%

%%%%%%%%%%%%%%%%%%%%%%%%%%%%%%%%%%%%%%%%%%%%%%%%%%%%%%%%%%%%%%%%%%%%%%%
\subsection{Inflationary Background}
\label{sec:infdynamics}
%%%%%%%%%%%%%%%%%%%%%%%%%%%%%%%%%%%%%%%%%%%%%%%%%%%%%%%%%%%%%%%%%%%%%%%
We assume a spatially flat Friedmann-Lema\^itre-Robertson-Walker (FLRW) background with line element
\begin{equation}
    ds^2 \; = \; a(\eta)^2\left(-d\eta^2 + d\mathbf{x}^2\right) \, ,
    \label{eq:frw_metric}
\end{equation}
where $\eta$ is conformal time ($d\eta \equiv dt/a$), $a(\eta)$ is the scale factor, $\mathbf{x}$ is the comoving spatial coordinate, and primes (dots) denote derivatives with respect to $\eta$ ($t$). The inflaton sector is described by the Einstein-Hilbert term plus a canonical scalar field,
\begin{equation}
    \mathcal{S}_\phi \; = \; \int d^4x\,\sqrt{-g}\, \left[
        \frac{M_P^2}{2}\, R
        -\frac{1}{2}(\partial_\mu\phi)^2
        -V(\phi)
         \right] \, ,
    \label{eq:infaction}
\end{equation}
with reduced Planck mass $M_P \equiv (8\pi G_N)^{-1/2} \simeq 2.435\times 10^{18}\,{\rm GeV}$.\footnote{We adopt the $(-, +, +, +)$ metric signature and the Riemann tensor convention $R^\mu{}_{\nu\rho\sigma} = \partial_\rho\Gamma^\mu_{\nu\sigma} - \cdots$, so that $R = 12H^2$ in de Sitter space.} Here $R$ is the Ricci scalar and $V(\phi)$ is the inflaton potential. Varying Eq.~\eqref{eq:infaction} yields the background inflaton equation of motion
\begin{equation}
    \phi'' + 2\mathcal{H}\phi' + a^2 V_{,\phi} \; = \; 0 \,,
    \label{eq:kg_conf}
\end{equation}
where $\mathcal{H}\equiv a'/a$ is the conformal Hubble rate and $V_{,\phi}\equiv dV/d\phi$. The expansion is governed by the Friedmann equation,
\begin{equation}
    \mathcal{H}^2 \;=\; (aH)^2 \;=\; \frac{a^2}{3M_P^2}\,\rho_\phi \,,
    \label{eq:friedmann_conf}
\end{equation}
where $H=\dot{a}/a$ denotes the Hubble parameter, and where the inflaton energy density is
\begin{equation}
    \rho_\phi \;=\; \frac{\phi'^2}{2a^2} + V(\phi) \, .
    \label{eq:rho_phi}
\end{equation}
We introduce the potential slow-roll parameters~\cite{Lyth:1998xn,Baumann:2009ds}
\begin{equation}
    \varepsilon_V \equiv \frac{M_P^2}{2}\left(\frac{V_{,\phi}}{V}\right)^2,
    \qquad
    \eta_V \equiv M_P^2\frac{V_{,\phi\phi}}{V},
    \label{eq:sr_params}
\end{equation}
which satisfy $\varepsilon_V,|\eta_V|\ll 1$ during slow roll. The number of $e$-folds between horizon exit of a comoving scale $k$ and the end of inflation is
\begin{equation}
    N(k)\;\equiv\;\ln\!\left(\frac{a_{\rm end}}{a_k}\right)
    \;\simeq\;
    \frac{1}{M_P^2}\int_{\phi_{\rm end}}^{\phi_k}\frac{V}{V_{,\phi}}\,d\phi
    \;=\;
    \int_{\phi_{\rm end}}^{\phi_k}\frac{d\phi}{M_P\sqrt{2\varepsilon_V}}\,.
    \label{eq:efolds}
\end{equation}
We adopt the pivot scale $k_* = 0.05~{\rm Mpc}^{-1}$ following \textit{Planck}~\cite{Planck:2018jri}, and denote by $\phi_*$ the inflaton field value at $k_*$-exit.

The end of inflation is defined by $\varepsilon \equiv -\dot H/H^2 = 1$, or equivalently $d(1/aH)/d\eta = 0$, when the comoving Hubble radius reaches its minimum.  For a canonical single field, the Friedmann equations imply $\dot\phi^2 = V(\phi)$ at $\varepsilon = 1$, giving
\begin{equation}
    \rho_{\rm end} \;\equiv\;\rho_\phi(a_{\rm end})
    \; = \;\frac{3}{2}\,V(\phi_{\rm end}),
    \qquad
    H_{\rm end}^2 \; = \;\frac{V(\phi_{\rm end})}{2M_P^2} \,.
    \label{eq:end_inflation_defs}
\end{equation}
Current CMB data tightly constrain the inflationary parameter space. BICEP/\textit{Keck} reports $r<0.036$ (95\% C.L.)~\cite{BICEP:2021xfz} while combined analyses yield the scalar spectral tilt $n_s\simeq 0.965\pm {\cal O}(10^{-3})$~\cite{Planck:2018jri,Tristram:2021tvh}. These benchmarks favor plateau-like potentials, including Starobinsky inflation~\cite{Starobinsky:1980te,Ellis:2013nxa} and $\alpha$-attractor models such as the T-model~\cite{Kallosh:2013hoa,Kallosh:2013maa}.

For concreteness in our explicit calculations we adopt the quadratic T-model potential
\begin{equation}
    V(\phi)
    \;=\;
    \lambda M_P^4
    \left[\sqrt{6}\tanh\!\left(\frac{\phi}{\sqrt{6}\,M_P}\right)\right]^2 \,,
    \label{eq:tmodel_potential}
\end{equation}
while emphasizing that our results extend to a broad class of plateau potentials with a quadratic minimum near the origin.\footnote{For a recent discussion of T-model constraints using ACT DR6 and SPT-3G data, see Ref.~\cite{Ellis:2025zrf}.}
Near $\phi=0$ the potential reduces to $V\simeq \frac12 m_\phi^2\phi^2$, with
\begin{equation}
    m_\phi \;=\; \sqrt{2\lambda}\,M_P.
    \label{eq:inflaton_mass}
\end{equation}
The normalization $\lambda$ is fixed by the amplitude of the primordial curvature spectrum at the pivot scale,
$\Delta_\zeta^2(k_*) \simeq 2.1\times 10^{-9}$~\cite{Planck:2018jri}. For the T-model, a convenient approximation in the large-$N_*$ limit is~\cite{Garcia:2020wiy,Ellis:2021kad}
\begin{equation}
    \lambda
    \;\simeq\;
    \frac{3\pi^2\,\Delta_\zeta^2(k_*)}{N_*^2}
    \;\simeq\;
    \frac{6.2\times 10^{-8}}{N_*^2} \,,
    \label{eq:lambda_norm}
\end{equation}
where $N_*\equiv N(k_*)$ is the number of $e$-folds between $k_*$-exit and the end of inflation.
Throughout we take $N_*=55$, noting that our main conclusions are only weakly sensitive to this choice.
For $N_*=55$ one obtains the standard attractor predictions~\cite{Kallosh:2013yoa}
\begin{equation}
    n_s \;\simeq\; 1 - \frac{2}{N_*} \;\simeq\; 0.964 \,,
    \qquad
    r \;\simeq\; \frac{12}{N_*^2} \;\simeq\; 0.004 \,,
    \label{eq:ns_r_tmodel}
\end{equation}
in good agreement with \textit{Planck} and BICEP/\textit{Keck}~\cite{Planck:2018jri,BICEP:2021xfz,Ellis:2021kad}.
Numerically, these values correspond to
\begin{equation}
    m_\phi \;\simeq\; 1.6\times 10^{13}\,{\rm GeV}\,,
    \qquad
    H_* \;\simeq\; 1.6\times 10^{13}\,{\rm GeV} \,,
    \qquad
    H_{\rm end}\;\simeq\; 6.3\times 10^{12}\,{\rm GeV} \,,
    \label{eq:scales_summary}
\end{equation}
with $\phi_{\rm end}\simeq 0.84 M_P$ for the end of inflation condition adopted above.\footnote{When quoting $H_I$ we mean the Hubble scale at $k_*$-exit, and we compute it consistently from
$H^2\simeq V/(3M_P^2)$ during slow roll.}

%%%%%%%%%%%%%%%%%%%%%%%%%%%%%%%%%%%%%%%%%%%%%%%%%%%%%%%%%%%%%%%%%%%%%%%
\subsection{Reheating}
\label{sec:reheating}
%%%%%%%%%%%%%%%%%%%%%%%%%%%%%%%%%%%%%%%%%%%%%%%%%%%%%%%%%%%%%%%%%%%%%%%
Inflation ends when the slow roll approximation breaks down and the inflaton begins coherent oscillations about the minimum of $V(\phi)$. The subsequent transfer of energy from the inflaton condensate to lighter degrees of freedom reheats the Universe and establishes the hot thermal bath required for Big Bang nucleosynthesis~\cite{Kawasaki:2000en,deSalas:2015glj,Hasegawa:2019jsa}.

For the T-model potential in Eq.~\eqref{eq:tmodel_potential}, the inflaton oscillates around an approximately quadratic minimum, so that the homogeneous condensate behaves as pressureless matter when averaging over the oscillation period, $\langle w_\phi\rangle \simeq 0$. We assume that the inflaton decays perturbatively to Standard Model particles with a constant rate $\Gamma_\phi$, and that collective effects associated with broad parametric resonance (``preheating'') can be neglected.\footnote{For the T-model, this assumption is satisfied, for example, for two-body decays into fermions with Yukawa coupling $y\lesssim 10^{-5}$~\cite{Garcia:2021iag}, or for multi-body decays as realized in no-scale inflation constructions~\cite{Ellis:2015kqa,Ellis:2020lnc}. Even when preheating is initially efficient, the background expansion history is well approximated by perturbative reheating after the first few oscillation cycles.} The background energy transfer is then captured by the coupled system
\begin{align}
    \rho_\phi' + 3\mathcal{H}\rho_\phi \;&=\; -a\,\Gamma_\phi\,\rho_\phi \, ,
    \label{eq:boltz_phi}\\
    \rho_R' + 4\mathcal{H}\rho_R \;&=\; a\,\Gamma_\phi\,\rho_\phi \,,
    \label{eq:boltz_rad}
\end{align}
supplemented by the Friedmann constraint
\begin{equation}
    3M_P^2\mathcal{H}^2 \;=\; a^2\left(\rho_\phi+\rho_R\right).
    \label{eq:friedreh}
\end{equation}
Here $\rho_R$ is the energy density of the relativistic plasma produced by inflaton decays.  We define the completion of reheating by the crossover time $t_{\rm reh}$ at which radiation becomes the dominant component,
\begin{equation}
    \rho_\phi(t_{\rm reh})=\rho_R(t_{\rm reh})\equiv \rho_{\rm reh}\,.
    \label{eq:reh_def}
\end{equation}
Assuming rapid thermalization of the decay products, we define the reheating temperature through the thermodynamic relation
\begin{equation}
    \rho_{\rm reh}
    \;=\;
    \frac{\pi^2}{30}\,g_{\rm reh}\,T_{\rm reh}^4\,,
    \label{eq:reh_temp_def}
\end{equation}
where $g_{\rm reh}$ is the effective number of relativistic degrees of freedom at reheating.  Throughout we take $g_{\rm reh}=427/4$, appropriate for the Standard Model at $T_{\rm reh}\gtrsim m_t$.\footnote{This crossover definition of reheating differs from the alternatively used condition $\Gamma_\phi = H$ by an $\mathcal{O}(1)$ factor in $T_{\rm reh}$.}

Before the crossover the inflaton condensate redshifts as matter, $\rho_\phi\propto a^{-3}$, and the total effective equation of state is $w_{\rm eff}\simeq 0$. After the crossover, $w_{\rm eff}\to 1/3$.  Solving Eqs.~\eqref{eq:boltz_phi}-\eqref{eq:friedreh} in the early matter-dominated regime yields~\cite{Giudice:2000ex,Garcia:2020eof}
\begin{equation}
    T_{\rm reh}
    \;\simeq\;
    \left(\frac{72}{5\pi^2 g_{\rm reh}}\right)^{1/4}
    \!\left(\Gamma_\phi M_P\right)^{1/2},
    \label{eq:TrehP}
\end{equation}
which shows that $T_{\rm reh}$ (or equivalently $\Gamma_\phi$) is the single free parameter encoding the reheating history. 
We define the comoving scale that re-enters the Hubble radius at the end of reheating, $k_{\rm reh} \;\equiv\; a_{\rm reh}\,H_{\rm reh}$, which marks the boundary between modes that re-enter the Hubble radius
during the matter-dominated reheating phase ($k\gg k_{\rm reh}$) and those whose sub-horizon evolution is entirely radiation-dominated
($k\ll k_{\rm reh}$). The latter experience a different transfer function for both scalar and tensor perturbations, imprinting the reheating history onto the gravitational wave spectrum.

In the remainder of the paper we parametrize the inflaton sector microphysics by specifying $T_{\rm reh}$, and we consistently evolve the full Boltzmann system~\eqref{eq:boltz_phi}-\eqref{eq:friedreh} to track the equation of state transition from $w_{\rm eff}\simeq 0$ to $w_{\rm eff}= 1/3$ and its impact on the spectator field perturbations and induced gravitational wave spectrum.

%%%%%%%%%%%%%%%%%%%%%%%%%%%%%%%%%%%%%%%%%%%%%%%%%%%%%%%%%%%%%%%%%%%%%%%
\subsection{Spectator Field and Effective Mass}
\label{sec:spectator}
%%%%%%%%%%%%%%%%%%%%%%%%%%%%%%%%%%%%%%%%%%%%%%%%%%%%%%%%%%%%%%%%%%%%%%%
We consider a spectator scalar field $\chi$ that is energetically subdominant during inflation. Our focus is on how the quantum fluctuations of $\chi$ generated during inflation, and subsequently processed through reheating, source a stochastic gravitational wave background at second order in perturbation theory. This setup encompasses a broad class of spectator candidates, including curvaton-like fields that decay after inflation~\cite{Linde:1996gt,Enqvist:2001zp,Moroi:2001ct,Lyth:2001nq} and stable relics produced gravitationally or via feeble interactions~\cite{Chung:1998zb,Chung:1998rq,Ling:2021zlj,Garcia:2022vwm,Kolb:2023ydq}. Related realizations include the SM Higgs as a  spectator~\cite{Lu:2019tjj,Litsa:2020mvj,Karam:2021qgn}, Peccei-Quinn fields~\cite{Peccei:1977ur,Peccei:1977hh}, and the scalar fields in supersymmetry/supergravity~\cite{Martin:1997ns,Freedman:2012zz}. Here we concentrate on the curvaton and dark matter interpretations, which cleanly connect the spectator fluctuation spectrum to late-time matter perturbations.

We take $\chi$ to be minimally coupled to gravity ($\xi=0$), with action given by Eq.~(\ref{eq:generalaction}), so that $\chi$ couples to the inflaton condensate through the portal $\frac12\sigma\phi^2\chi^2$ and has a quartic self-interaction $\lambda_\chi\chi^4/4!$.\footnote{A nonminimal coupling $\frac12\xi R\chi^2$ can substantially modify both the inflationary and post-inflationary evolution, including the effective mass and the resonance structure~\cite{Garcia:2023qab, Verner:2024agh}. We leave a systematic exploration of $\xi\neq 0$ to future work.} Varying Eq.~\eqref{eq:generalaction} gives the spectator equation of motion,
\begin{equation}
    \chi'' + 2\mathcal{H}\chi' - \nabla^2\chi
    + a^2\!\left(m_\chi^2 + \sigma\phi^2\right)\chi
    + a^2\,\frac{\lambda_\chi}{6}\,\chi^3
    \;=\; 0\,.
    \label{eq:eom_chi_full}
\end{equation}
Introducing the rescaled field $X(\eta,\mathbf{x}) \;\equiv\; a(\eta)\,\chi(\eta,\mathbf{x})$ absorbs the Hubble friction and yields
\begin{equation}
    X'' - \nabla^2 X
    + \left[a^2\!\left(m_\chi^2+\sigma\phi^2\right) - \frac{a''}{a}\right] X
    + \frac{\lambda_\chi}{6}\,X^3
    \;=\; 0\,.
    \label{eq:eom_X_full}
\end{equation}
In the Gaussian/Hartree approximation (linear mode equation with a self-consistent effective mass), Eq.~\eqref{eq:eom_X_full} reduces to
\begin{equation}
    \left[
        \frac{d^2}{d\eta^2}
        -\nabla^2
        + a^2 m_{\chi,\rm eff}^2(\eta)
    \right]X(\eta,\mathbf{x})
    \;=\; 0\,,
    \label{eq:eom_X_linear}
\end{equation}
where the time-dependent effective mass-squared is
\begin{equation}
    m_{\chi,\rm eff}^2(\eta)
    \;\equiv\;
    m_\chi^2+\sigma\,\phi(\eta)^2
    -\frac{R}{6}
    +\frac{\lambda_\chi}{2}\,\langle\chi^2\rangle\,.
    \label{eq:meff_total}
\end{equation}
The first two terms are the bare mass and the inflaton portal. The $-R/6$ contribution arises from absorbing the $-a''/a$ factor in Eq.~\eqref{eq:eom_X_full} via the identity $a''/a = a^2 R/6$ for a spatially flat FLRW background. Finally, the Hartree term $\frac{\lambda_\chi}{2}\langle\chi^2\rangle$ captures the leading mean-field backreaction of the quartic self-interaction on the mode evolution. It is obtained by replacing $\chi^3\to 3\langle \chi^2\rangle \chi$ in Eq.~\eqref{eq:eom_chi_full}.

During quasi-de Sitter inflation one has $R\simeq 12H^2$ (up to
slow-roll corrections), so the $-a''/a$
term in Eq.~\eqref{eq:eom_X_full} contributes $-a^2R/6\simeq -2(aH)^2$
to the effective frequency of $X=a\chi$.  This contribution is a purely
geometric consequence of the field redefinition: for a massless
minimally coupled spectator, the physical field $\chi_k$ simply freezes
on superhorizon scales, and the apparent growth of $X_k\propto a$
merely reflects the prefactor.  The physical superhorizon evolution is
instead controlled by
\begin{equation}
    M^2(\eta)
    \;\equiv\;
    m_\chi^2+\sigma\,\phi(\eta)^2
    +\frac{\lambda_\chi}{2}\,\langle\chi^2\rangle\,,
    \label{eq:Msq_def}
\end{equation}
which collects the non-gravitational contributions to the spectator
effective mass. A positive $M^2$ causes superhorizon modes to decay,
producing a blue-tilted spectrum that is suppressed on large scales. A vanishing $M^2$ gives (nearly) scale-invariant fluctuations in quasi-de Sitter (and exact scale invariance in pure de Sitter).  The spectral tilt is given by $n_\chi-1\simeq 2M^2/(3H_*^2)$ for $M^2\ll H_*^2$ (up to slow-roll corrections), and since all
contributions in Eq.~\eqref{eq:Msq_def} are non-negative, the spectrum
is always blue-tilted or scale-invariant for $\xi=0$. This is the key
feature exploited in this work: large spectator fluctuations on small
scales can source a detectable GW signal while remaining consistent
with the stringent CMB isocurvature bound on large
scales~\cite{Chung:2004nh,Ling:2021zlj,Garcia:2023awt,Ebadi:2023xhq,Garcia:2025yit}.

%%%%%%%%%%%%%%%%%%%%%%%%%%%%%%%%%%%%%%%%%%
\section{Spectator Particle Production}
\label{sec:gpp}
%%%%%%%%%%%%%%%%%%%%%%%%%%%%%%%%%%%%%%%%%%
A central ingredient in the GW prediction is the spectator abundance
$\rho_\chi(\eta)$ and the associated mode-function evolution
$X_k(\eta)$, which together determine the amplitude and spectral
shape of the scalar-induced GW signal. In the parameter regime of
interest, i.e., large portal coupling $\sigma\phi^2\chi^2$ with
$\sigma/\lambda\gg 1$, the dominant production mechanism is
\emph{parametric amplification} of spectator modes by the
oscillating inflaton condensate during
reheating~\cite{Kofman:1994rk,Shtanov:1994ce,Kofman:1997yn}.  This
is qualitatively distinct from the purely gravitational production
that operates for $\sigma\to 0$: whereas gravitational production
creates at most $\mathcal{O}(1)$ quanta per mode at the
inflation-reheating transition, parametric resonance can amplify the
occupation number by many orders of magnitude over a few $e$-folds,
building up the large spectator power spectrum required for an
observable GW signal.

A purely gravitationally produced spectator ($\sigma=0$) yields a
relic abundance that is too small to generate a detectable GW
spectrum for the reheating temperatures consistent with dark matter
overproduction constraints~\cite{Garcia:2022vwm,Garcia:2023qab,Garcia:2025rut}.
The large portal coupling simultaneously solves both problems: it
exponentially enhances the spectator power spectrum through
parametric resonance while suppressing the superhorizon fluctuations
during inflation (due to the heavy effective mass
$\sigma\phi_*^2\gg H_*^2$), thereby relaxing the isocurvature
bound~\cite{Garcia:2023dyf,Garcia:2025wmu}.

%%%%%%%%%%%%%%%%%%%%%%%%%%%%%%%%%%%%%%%%%%
\subsection{Mode Equations and Bogoliubov Formalism}
%%%%%%%%%%%%%%%%%%%%%%%%%%%%%%%%%%%%%%%%%%
We work with the rescaled field $X\equiv a\chi$. Quantizing in the
FLRW background, we expand the field operator as
\begin{equation}
\label{eq:xfourier}
    X(\eta,\mathbf{x})
    \;=\;
    \int\frac{d^3k}{(2\pi)^{3/2}}\,
    e^{-i\mathbf{k}\cdot\mathbf{x}}
    \left[X_k(\eta)\,\hat a_\mathbf{k}
    +X_k^*(\eta)\,\hat a_{-\mathbf{k}}^\dagger\right],
\end{equation}
where $\mathbf{k}$ is the comoving momentum and
$[\hat a_\mathbf{k},\hat a_{\mathbf{k}'}^\dagger]
=\delta^{(3)}(\mathbf{k}-\mathbf{k}')$.
The mode functions obey
\begin{equation}
\label{eq:eomX}
    X_k''(\eta)+\omega_k^2(\eta)\,X_k(\eta)=0\,,
\end{equation}
with time-dependent frequency
\begin{equation}
\label{eq:omegak}
    \omega_k^2(\eta)
    \;=\;
    k^2+a^2(\eta)\,m_{\chi,\rm eff}^2(\eta)\,,
\end{equation}
where $m_{\chi,\rm eff}^2(\eta)$ is the effective spectator mass
defined in Eq.~\eqref{eq:meff_total}, including the bare mass, the
inflaton portal $\sigma\phi^2$, the curvature contribution $-R/6$,
and the leading Hartree correction from the quartic
self-interaction. Canonical quantization requires the Wronskian
normalization
\begin{equation}
\label{eq:wronskian}
    X_k X_k^{\prime*}-X_k^*X_k'=i\,,
\end{equation}
which is preserved by Eq.~\eqref{eq:eomX} and guarantees the
equal time canonical commutation relations.

The particle content of each mode is characterized by the Bogoliubov
coefficients $\alpha_k(\eta)$ and $\beta_k(\eta)$, defined through
the instantaneous projections of the exact solution onto the
positive- and negative-frequency WKB
basis~\cite{Parker:1969au,Birrell:1982ix}:
\begin{equation}
\label{eq:bogo_def}
    \alpha_k\,f_k
    \;\equiv\;
    \tfrac{1}{2}\!\left(X_k+\frac{i}{\omega_k}X_k'\right),
    \qquad
    \beta_k\,f_k^*
    \;\equiv\;
    \tfrac{1}{2}\!\left(X_k-\frac{i}{\omega_k}X_k'\right),
\end{equation}
where
\begin{equation}
\label{eq:fk_WKB}
    f_k(\eta)
    \;\equiv\;
    \frac{1}{\sqrt{2\omega_k(\eta)}}\,
    \exp\!\left(-i\!\int^\eta\!\omega_k(\tilde\eta)\,
    d\tilde\eta\right).
\end{equation}
By construction,
$X_k=\alpha_k f_k+\beta_k f_k^*$, and the Wronskian
condition~\eqref{eq:wronskian} implies the unitarity
relation\footnote{The instantaneous-projection
  definition~\eqref{eq:bogo_def} differs from the slowly-varying
  (constrained) Bogoliubov coefficients used in some
  references (see e.g.~\cite{Kofman:1997yn}) by terms of order
  $\omega_k'/\omega_k^2$, which vanish in the adiabatic limit.  Both
  conventions yield the same physical particle number once production
  has ceased and the evolution becomes adiabatic.}
\begin{equation}
\label{eq:unitarity}
    |\alpha_k|^2-|\beta_k|^2=1\,.
\end{equation}
The occupation number of mode $k$ is
\begin{equation}
\label{eq:nk_def}
    n_k(\eta)
    \;\equiv\;
    |\beta_k(\eta)|^2
    \;=\;
    \frac{1}{2\omega_k(\eta)}
    \left|\omega_k(\eta)\,X_k(\eta)-iX_k'(\eta)\right|^2.
\end{equation}
In an adiabatic regime, $n_k$ is conserved. Non-adiabatic evolution
of $\omega_k$ drives $n_k$ away from zero, producing a non-thermal
spectator population.

%%%%%%%%%%%%%%%%%%%%%%%%%%%%%%%%%%%%%%%%%%
\subsection{Parametric Amplification During Reheating}
%%%%%%%%%%%%%%%%%%%%%%%%%%%%%%%%%%%%%%%%%%
In the parameter regime of interest, spectator particle production
proceeds through three distinct phases:

(i)~\emph{During inflation}, the large portal coupling keeps the
effective mass heavy, $\sigma\phi_*^2\gg H_*^2$, placing the
spectator deep in the non-adiabatic suppression regime.  Superhorizon
modes are exponentially damped rather than frozen, yielding an
extremely suppressed spectrum at the end of inflation (see the
$N-N_{\rm end}=0$ curve in Fig.~\ref{fig:PX_T}).

(ii)~\emph{At the end of inflation}, the rapid change in the
background equation of state produces a burst of non-adiabatic
particle creation, seeding a small but nonzero occupation number in
modes with $k\lesssim a_{\rm end}m_\phi$.

(iii)~\emph{During reheating}, the oscillating inflaton condensate
$\phi(\eta)\simeq\phi_{\rm end}(a_{\rm end}/a)^{3/2}\cos(m_\phi t)$
drives a periodically varying effective mass
$m_{\chi,\rm eff}^2\supset\sigma\phi^2(\eta)$.  This induces
parametric resonance in the spectator mode
equation~\eqref{eq:eomX}: every half-period of the inflaton
oscillation, $\omega_k^2$ passes through a non-adiabatic regime,
producing a burst of amplification.  For large $\sigma$, the resonance
parameter $q\equiv\sigma\phi_{\rm end}^2/(4m_\phi^2)\gg 1$, placing
the system in the broad resonance regime where exponential growth of
$n_k$ can persist over many inflaton
oscillations~\cite{Kofman:1994rk,Kofman:1997yn,Shtanov:1994ce}.
The process is analogous to
preheating~\cite{Kofman:1997yn,Shtanov:1994ce}, but here it
populates the spectator sector rather than the visible sector.
Lattice simulations have demonstrated that such parametric
amplification through a quadratic portal coupling produces a
significant stochastic GW background, with amplitude and spectral
shape controlled by the resonance
parameter~\cite{Figueroa:2017vfa,Cosme:2022htl,Cui:2023fbg}, and
have quantified the role of fragmentation and backreaction beyond the
Hartree
approximation~\cite{Garcia:2023qab,Bettoni:2024ixe}.

The exponential amplification of $n_k$ is eventually regulated by the
quartic self-interaction $\lambda_\chi\chi^4/4!$, which generates a
Hartree mass $\frac{\lambda_\chi}{2}\langle\chi^2\rangle$ that grows
with the spectator variance and eventually detunes the resonance.
This backreaction mechanism is captured self-consistently in our
Hartree approximation, where $\langle\chi^2\rangle$ enters the
effective mass~\eqref{eq:meff_total} and is updated at each time
step.  The competition between parametric driving (controlled by
$\sigma$) and Hartree backreaction (controlled by $\lambda_\chi$)
determines the final spectator abundance, with the ratio
$\sigma/\lambda_\chi$ being the key parameter governing the
amplification efficiency~\cite{Garcia:2022vwm,Garcia:2023qab,Garcia:2025rut}.
Once the inflaton amplitude has decayed sufficiently (through Hubble
friction and perturbative decay) for the spectator evolution to
become adiabatic, $n_k$ freezes to a final value $n_k^{(\rm f)}$.
In the non-relativistic regime ($m_\chi\gg H$), $\rho_\chi$
subsequently redshifts as matter ($\propto a^{-3}$).

The vacuum-subtracted physical energy density of the produced
spectator quanta can be written in compact normal-ordered
form~\cite{Chung:2004nh,Kolb:2023ydq}:\footnote{A detailed
  derivation of Eq.~\eqref{eq:rho_chi}, including the
  normal-ordering prescription in an expanding background, is given
  in Appendix~A of Ref.~\cite{Garcia:2025wmu}.  That $\rho_\chi$ is
  indeed vacuum-subtracted can be seen by expanding the square and
  using the Wronskian~\eqref{eq:wronskian}:
  $|\omega_kX_k-iX_k'|^2=\omega_k^2|X_k|^2+|X_k'|^2-\omega_k$,
  where the $-\omega_k$ removes the zero-point contribution.
  Equivalently, by the definition of
  $n_k$~\eqref{eq:nk_def}, the integrand equals $2\omega_k n_k$,
  confirming that only produced quanta contribute.}
\begin{equation}
\label{eq:rho_chi}
    \rho_\chi(\eta)
    \;=\;
    \frac{1}{2a^4}\int\frac{d^3k}{(2\pi)^3}
    \left|\omega_k(\eta)\,X_k(\eta)-iX_k'(\eta)\right|^2
    \;=\;
    \frac{1}{a^4}\int\frac{d^3k}{(2\pi)^3}\,
    \omega_k(\eta)\,n_k(\eta)\,,
\end{equation}
making explicit that $\omega_k/a$ is the physical energy per quantum.
For later use we introduce the comoving number density and energy
density per logarithmic momentum interval,
\begin{equation}
\label{eq:Nk_def}
    n_\chi(\eta)\,a^3
    \;=\;
    \int_{k_0}^\infty\frac{dk}{k}\,\mathcal{N}_k(\eta)\,,
    \qquad
    \mathcal{N}_k(\eta)
    \;\equiv\;
    \frac{k^3}{2\pi^2}\,n_k(\eta)\,,
\end{equation}
and
\begin{equation}
\label{eq:Ek_def}
    \rho_\chi(\eta)\,a^3
    \;=\;
    \int_{k_0}^\infty\frac{dk}{k}\,\mathcal{E}_k(\eta)\,,
    \qquad
    \mathcal{E}_k(\eta)
    \;\equiv\;
    \frac{k^3}{2\pi^2}\,\frac{\omega_k(\eta)}{a(\eta)}\,
    n_k(\eta)\,.
\end{equation}
Here $k_0=a_0H_0$ is an infrared cutoff set by the present-day
Hubble radius, while the ultraviolet behavior is regulated by the
rapid suppression of $n_k$ once the mode evolution becomes
adiabatic.

We initialize the mode functions in the Bunch-Davies vacuum deep
inside the horizon during inflation, at an initial conformal time $\eta_0$ satisfying
$|k\eta_0|\gg 1$ for all modes of interest:
\begin{equation}
\label{eq:BD_init}
    X_k(\eta_0)=\frac{1}{\sqrt{2\omega_k(\eta_0)}}\,,
    \qquad
    X_k'(\eta_0)=-i\,\omega_k(\eta_0)\,X_k(\eta_0)\,.
\end{equation}
We then evolve Eq.~\eqref{eq:eomX} across the end of inflation,
through reheating (using the background described in
Sec.~\ref{sec:reheating}), and into the radiation-dominated era.
The Hartree term $\frac{\lambda_\chi}{2}\langle\chi^2\rangle$ in
$m_{\chi,\rm eff}^2$ is evaluated self-consistently at each time
step using the renormalized variance~\eqref{eq:X2_integral}. This procedure captures the full
non-perturbative dynamics of the spectator sector, including parametric amplification, Hartree backreaction, and the eventual freeze-out of the occupation number, providing the mode functions
$X_k(\eta)$ and the abundance history $\rho_\chi(\eta)$ required for
the GW computation in subsequent sections. As a cross check on the Hartree approximation, the resulting power spectrum is compared to full nonlinear lattice simulations using \textsc{CosmoLattice}~\cite{Figueroa:2020rrl,Figueroa:2021yhd} (see Figs.~\ref{fig:rhos_sl} and~\ref{fig:PX_T}). The
excellent agreement for $k\lesssim k_{\rm end}$, where the bulk of
the GW signal originates, validates the mean-field treatment for our
purposes.

\begin{figure*}[t!]
    \centering
    \includegraphics[width=\linewidth]{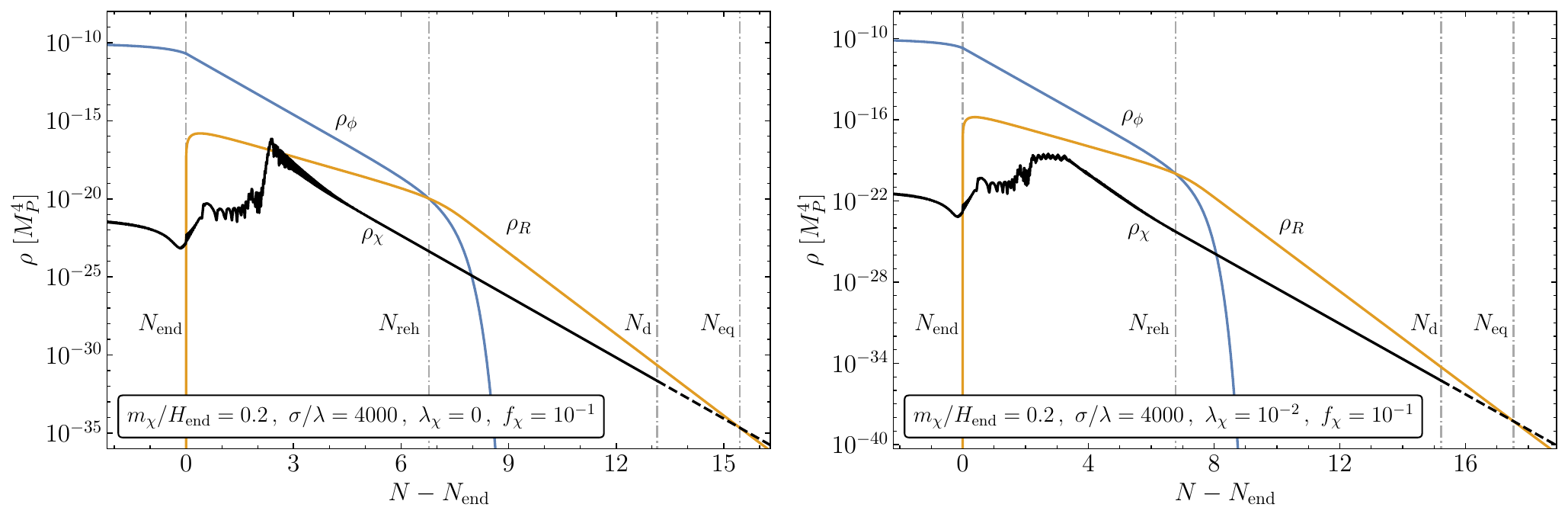}
    \caption{Energy densities of the inflaton, $\rho_{\phi}$ (blue), the visible sector radiation, $\rho_R$ (orange), and the spectator scalar, $\rho_{\chi}$ (black), as functions of the number of $e$-folds, for a high reheating temperature, $T_{\rm reh}=10^{13}\ {\rm GeV}$ and a large spectator-inflaton coupling. The continuous black curve follows $\rho_{\chi}$ until its decay, which is assumed to occur when $f_{\chi}=\rho_{\chi}/\rho_R=10^{-1}$. The dashed line follows $\rho_{\chi}$ in the absence of a decay. \textbf{Left:} Evolution in the absence of a spectator self-interaction. \textbf{Right:} Evolution in the presence of $\chi$ self-interaction, with $\lambda_{\chi}=10^{-2}$.}
    \label{fig:rhos}
\end{figure*}

%%%%%%%%%%%%%%%%%%%%%%%%%%%%%%%%%%%%%%%%%%
\subsection{Evolution Through Reheating}
%%%%%%%%%%%%%%%%%%%%%%%%%%%%%%%%%%%%%%%%%%
Fig.~\ref{fig:rhos} shows in detail the evolution of the three sectors that encompass our cosmological scenario. The blue curve, which dominates the energy density before the end of inflation (at $N=N_{\rm end}$ $e$-folds) and the end of reheating (at $N=N_{\rm reh}$) is the inflaton energy density $\rho_{\phi}$. The energy density of the visible sector radiation, $\rho_R$, is shown as the orange curve. As stated above, we assume its production can be described by perturbative means, allowing us to disregard the presence of a non-vanishing $\rho_R$ during inflation. Upon solution of the system (\ref{eq:boltz_phi})-(\ref{eq:friedreh}), a rapid growth to a maximum value for this energy component is found within the first oscillation of the inflaton. Afterwards, competition between particle production and dilution implies that $\rho_R\propto a^{-3/2}$ during reheating~\cite{Garcia:2020eof}. After reheating ends, $\rho_R\propto a^{-1}$, dominating the universe until matter-radiation equality, which would occur at $N=N_{\rm eq}$ for a stable $\chi$. For the spectator $\chi$ we distinguish two scenarios: unstable (curvaton-like) and stable (dark matter-like) spectator field. For the unstable case, we assume that $\chi$ decays into radiation at the time when the fractional energy density 
\beq
f_{\chi} \;\equiv\; \frac{\rho_{\chi}}{\rho_R} \;\leq\;1\,.
\eeq
In Fig.~\ref{fig:rhos} these two scenarios for the spectator $\chi$ are distinguished as the continuous black curve (unstable), which finishes at the decay time with $N=N_{\rm d}$, and the dashed black line (stable), which extends to matter-radiation equality and beyond. The left panel of Fig.~\ref{fig:rhos} shows the numerical integration of (\ref{eq:Ek_def}) for a spectator with a mass $m_{\chi}=0.2H_{\rm end}\simeq 0.08 m_{\phi}$, a sizeable coupling to $\phi$, $\sigma/\lambda=4000$, but no self-interaction, $\lambda_{\chi}=0$. As discussed above, for such a large coupling the growth of super-horizon modes during inflation is suppressed by the large effective mass, and the energy density $\rho_{\chi}$ is in consequence suppressed by over ten orders of magnitude relative to $\rho_{\phi}$ during inflation. Upon the start of reheating, the first burst of particle production can be clearly appreciated, as the field enters the parametric resonance regime, briefly overtaking the radiation energy density at the peak of the non-perturbative particle production. As expansion shuts down the resonance, the energy density rapidly decreases, until the adiabatic regime is reached, and the field redshifts like pressureless matter, $\rho_{\chi}\propto a^{-3}$. This behavior is maintained until $\chi$ decays at $f_{\chi}=0.1$ if unstable, or continues past $N_{\rm eq}$, if $\chi$ is stable. The right panel of Fig.~\ref{fig:rhos} shows the evolution of $\rho_{\chi}$ for a spectator of equal mass and inflaton coupling as in the left panel, but for a non-vanishing self-interaction of strength $\lambda_{\chi}=10^{-2}$. In this case, the suppression of the resonant growth can be immediately appreciated, as the energy density of the spectator lack a peak higher than $\rho_R$, leading to a smaller comoving energy density, which can be easily appreciated in the value of $N_{\rm eq}$.

The choice of a high reheating temperature for Fig.~\ref{fig:rhos} allows us to visualize the entirety of the evolution of the energy densities up to matter-radiation equality in less than 20 $e$-folds. In addition, as we will show explicitly in Section~\ref{sec:gwmat}, there is a direct correlation between the amplitude of the GW spectrum and the reheating temperature. For the nominal choice $T_{\rm reh}=10^{13}\,{\rm GeV}$, the reheating temperature is large, while being sufficiently low to allow the $\chi$ field to reach the adiabatic, non-relativistic regime before the end of reheating, greatly simplifying our discussion. A drawback of high $T_{\rm reh}$, however, is the impossibility to identify a stable $\chi$ with the present-day dark matter. Fig.~\ref{fig:rhos} shows clearly that the matter-radiation equality occurs very early, leading to an overabundance of $\chi$ at later times. More precisely, the relic abundance of $\chi$ can be estimated as
\beq
\Omega_{\chi}h^2 \;=\; \frac{\rho_{\chi}(a_0)}{\rho_c h^{-2}} \;=\; \frac{\rho_{\rm reh}}{\rho_c h^{-2}} \left(\frac{\rho_{\chi,{\rm reh}}}{\rho_{\rm reh}}\right)\left(\frac{a_{\rm reh}}{a_0}\right)^3\,,
\eeq
where $\rho_c$ denotes the critical energy density at the present time. Assuming a standard thermal history after the end of reheating, this expression can be rewritten as
\begin{align}
\Omega_{\rm DM}h^2 \;=\; \left(\frac{43 \pi^2}{330}\right) \frac{T_{\rm reh} T_0^3}{\rho_c h^{-2}} \left(\frac{\rho_{\chi,{\rm reh}}}{\rho_{\rm reh}}\right) \;\simeq\; \left(\frac{43 \pi^2}{990}\right) \frac{T_{\rm reh} T_0^3}{\rho_c h^{-2}} \left(\frac{\rho_{\chi,{\rm NR}}}{H^2_{\rm NR} M_P^2}\right)\,,
\end{align}
with $T_0=2.7255\,{\rm K}$ and $\rho_c h^{-2} = 8.1\times 10^{-47}\,{\rm GeV}^4$~\cite{Fixsen:2009ug,Planck:2018vyg}. Here we introduce the subscript NR to denote quantities evaluated at the onset of the adiabatic, non-relativistic regime for the (heavy) spectator field. For Fig.~\ref{fig:rhos} we can take $a_{\rm NR}\simeq 200$, which corresponds to $\rho_{\phi,{\rm NR}}\simeq 7.8\times 10^{-10}M_P^4$, and $\rho_{\chi,{\rm NR}}\simeq3.7\times 10^{-22} M_P^4$ for $\lambda_{\chi}=0$ ($\rho_{\chi,{\rm NR}}\simeq7.6\times 10^{-23} M_P^4$ for $\lambda_{\chi}=10^{-2}$). This would then require $T_{\rm reh}\simeq 1.2\,{\rm TeV}$ ($5.9\,{\rm TeV}$), far too low for GW detectability prospects in the near future. For this reason, we will focus our study of the associated GW spectra on the unstable, curvaton-like $\chi$ scenario.\par\medskip

\begin{figure*}[t!]
    \centering
    \includegraphics[width=\linewidth]{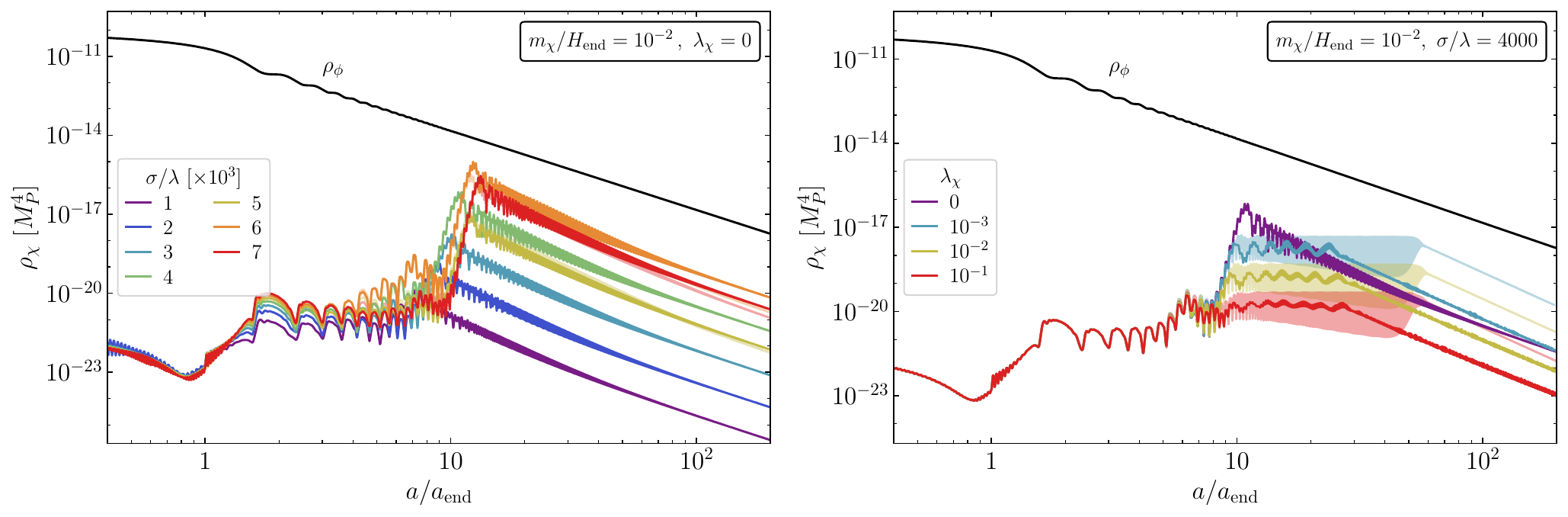}
    \caption{Scale factor dependence of the spectator field energy
    density $\rho_\chi$ during the early stages of reheating,
    compared to the inflaton energy density $\rho_\phi$ (black
    curve). \textbf{Left:} Effect of varying the
    inflaton-spectator coupling $\sigma/\lambda=(1$--$7)\times 10^3$ at fixed $m_\chi/H_{\rm end}=10^{-2}$ and $\lambda_\chi=0$.  The
    non-monotonic ordering of the final amplitudes at large
    $\sigma/\lambda$ reflects the complex structure of the broad
    parametric resonance bands. \textbf{Right:} Effect of varying
    the self-interaction $\lambda_\chi=0,\,10^{-3},\,10^{-2},\,
    10^{-1}$ at fixed $\sigma/\lambda=4000$ and
    $m_\chi/H_{\rm end}=10^{-2}$.  Increasing $\lambda_\chi$
    suppresses the resonant growth by generating a Hartree mass that
    detunes the resonance.  In both panels, fully opaque curves show
    the Hartree approximation, while translucent curves of the same
    color show the corresponding result from a lattice simulation
    with \textsc{CosmoLattice}. The two approaches agree well during the linear resonance phase. Deviations at late times arise from mode-mode rescattering and inflaton fragmentation effects beyond the mean-field
    approximation.}
    \label{fig:rhos_sl}
\end{figure*}

Fig.~\ref{fig:rhos_sl} shows the time evolution of the spectator energy density $\rho_\chi$ for a variety of couplings during the early stages of reheating. The left panel depicts the outcome of varying the inflaton-spectator coupling for a fixed mass $m_\chi/H_{\rm end}=10^{-2}$ and vanishing self-interaction. As expected, after the first burst of post-inflationary particle production, $\rho_\chi$ follows a monotonically growing relation with the coupling strength $\sigma$.  Nevertheless, as the inflaton oscillations accumulate, the presence of parametric resonance
becomes manifest. Owing to the nontrivial structure of the resonance bands, the monotonic dependence on $\sigma$ is lost, and in some cases increasing the coupling results in a decrease of the final comoving energy density (see e.g.~\cite{Garcia:2022vwm} for details).

Our derivation of the GW spectrum associated with the growth of the
$\chi$ fluctuations assumes that the evolution of the spectator mode function is independent, following the linear system~\eqref{eq:eomX}. In the presence of backreaction and rescatterings, mode-mode couplings induced by the fragmentation of the inflaton field can no longer be ignored, and the Hartree approximation breaks down~\cite{Lebedev:2021xey,Figueroa:2023oxc}. The dynamics of the system can still be tracked by means of lattice methods, but analytical control over the computation is lost. In the figure we show as fully opaque curves those computed in the Hartree approximation, and as translucent curves those determined from the lattice. We find that only for couplings close to the backreaction regime are noticeable deviations found. We refrain from considering larger couplings than those shown in Fig.~\ref{fig:rhos_sl}, for which the fragmentation of $\phi$ is, at worst, incomplete.

The right panel of Fig.~\ref{fig:rhos_sl} shows the evolution of
$\rho_\chi$ when the mass and portal coupling $\sigma$ are fixed but
the self-interaction $\lambda_\chi$ is varied.  Here we see the
expected curtailment of parametric growth: a larger $\lambda_\chi$
generates a stronger Hartree mass
$\frac{\lambda_\chi}{2}\langle\chi^2\rangle$ that detunes the
resonance earlier, leading to a weaker final
amplitude~\cite{Figueroa:2023oxc,Figueroa:2024asq}.  As
$\lambda_\chi$ is increased, the quartic term dominates the
potential energy of $\chi$, and the onset of the non-relativistic
regime is delayed accordingly.  A similar suppression of preheating
efficiency by the self-interaction was found in lattice studies of
Higgs-portal
reheating~\cite{Lebedev:2021xey}, where the Higgs quartic coupling
was shown to shut off the resonance, leaving only a small fraction
of the inflaton energy transferred to the daughter field.

The comparison between the Hartree approximation and the full
lattice computation reveals an important qualitative difference.
Until the onset of broad parametric resonance, both approaches yield
identical estimates of $\rho_\chi$.  However, after the largest
burst of particle production, mode-mode couplings become
non-negligible: on the lattice, the $\lambda_\chi\chi^4$ vertex
enables $2\to 2$ scattering processes that redistribute spectral
weight from the resonance band to shorter wavelengths, generating
additional UV power not captured by the Hartree
approximation~\cite{Micha:2002ey,Micha:2004bv,Figueroa:2023oxc}.
This turbulent cascade is visible in the right panel of
Fig.~\ref{fig:rhos_sl} as the enhanced late-time amplitude of the
translucent (lattice) curves relative to the opaque (Hartree) ones,
particularly for $\lambda_\chi=10^{-3}$. A systematic comparison between the quantum 2PI Hartree truncation and classical lattice simulations for a self-interacting spectator
was performed in Refs.~\cite{Kainulainen:2022lzp, Kainulainen:2024etd}, where
$\mathcal{O}(1)$ agreement in the final two-point function was
found, along with a novel nonlinear resonance driven by
$\langle\chi^2\rangle$ that can dominate the total particle yield.
Similar behavior has been observed in lattice studies of
non-minimally coupled spectators, where the self-interaction
regularizes the tachyonic growth while simultaneously feeding a
turbulent UV cascade~\cite{Bettoni:2021zhq,Figueroa:2024asq}.\footnote{The
discrepancy between Hartree and lattice results at nonzero $\lambda_\chi$ does not invalidate our GW predictions: the GW signal is dominated by modes near the spectral peak ($k\sim k_{\rm end}$), where the two approaches agree well. The UV tail, where the lattice shows enhanced power from rescattering, contributes subdominantly to $\Omega_{\rm GW}$.} This difference will be appreciable also in the corresponding GW
spectra, as we discuss in Section~\ref{sec:gwsignals}.\par\medskip

%%%%%%%%%%%%%%%%%%%%%%%%%%%%%%%%%%%%%%%%%%
\section{Spectator Field Power Spectrum}
\label{sec:matterPS}
%%%%%%%%%%%%%%%%%%%%%%%%%%%%%%%%%%%%%%%%%%
Having discussed the background evolution of the scalar spectator $\chi$, we now proceed to track the growth of its inhomogeneities, by means of its power spectrum. The scalar-induced GW signal computed in Section~\ref{sec:SIGWs} is sourced precisely by the matter power spectrum of the spectator sector. Here we define the relevant spectral quantities, discuss their evolution through reheating, and connect the field-level power spectrum to the density contrast that enters the GW source term.

We define the dimensionless, UV-subtracted power spectrum of the
rescaled field $X=a\chi$ as
\begin{equation}
    \Delta_X^2(k,\eta)
    \;\equiv\;
    \frac{k^3}{2\pi^2\,M_P^2}
    \left(|X_k(\eta)|^2 - \frac{1}{2\omega_k(\eta)}\right),
    \label{eq:DeltaX}
\end{equation}
so that the vacuum-subtracted variance is
\begin{equation}
    \frac{\langle X^2(\eta)\rangle}{M_P^2}
    \;=\;
    \int\frac{dk}{k}\,\Delta_X^2(k,\eta)\,.
    \label{eq:X2_integral}
\end{equation}
The subtraction of the instantaneous adiabatic vacuum $1/(2\omega_k)$
ensures ultraviolet convergence and connects to the Bogoliubov
formalism of Sec.~\ref{sec:gpp}: in the adiabatic regime,
$|X_k|^2\simeq(2n_k+1)/(2\omega_k)$, so the subtracted integrand
reduces to $n_k/\omega_k$, confirming that only the produced quanta
contribute~\cite{Birrell:1982ix,Chung:2004nh}.

After inflation, once the spectator becomes non-relativistic, each mode oscillates as
$\chi_k\propto a^{-3/2}$
$\cos(m_\chi t+\varphi_k)$. Time-averaging
over many oscillation cycles gives
$|X_k|^2=a^2|\chi_k|^2\propto a^{-1}$, so
$\Delta_X^2\propto a^{-1}$.  To isolate the spectral shape from this
trivial redshift, we introduce the rescaled spectrum
\begin{equation}
    \widetilde\Delta_X^2(k,\eta)
    \;\equiv\;
    \frac{a(\eta)}{a_{\rm end}}\,\Delta_X^2(k,\eta)\,,
    \label{eq:tildeDeltaX}
\end{equation}
which freezes to a time-independent function
$\widetilde\Delta_X^2(k,\eta)\to\widetilde\Delta_X^2(k)$ once all
modes of interest have become non-relativistic and particle production
has ceased. This frozen, vacuum-subtracted power spectrum $\widetilde\Delta_X^2(k)$, set by the interplay of parametric amplification during reheating and the self-consistent Hartree evolution of the effective mass, enters directly into the GW source term through the mode functions $X_k(\eta)$. In the next section we derive the master formula for the scalar-induced GW spectrum sourced by the field-gradient term $\partial_a\chi\,\partial_b\chi$.

For super-horizon modes ($k\ll k_{\rm end}\equiv a_{\rm end} H_{\rm end}$) in the large-coupling regime ($\sigma\phi_*^2\gg m_\chi^2,\,H_*^2$), the spatial gradient is
negligible and the effective frequency is $k$-independent, $\omega_k^2\simeq a^2(m_\chi^2+\sigma\phi^2)$. All long-wavelength
modes therefore evolve identically, giving $|X_k|^2$ independent of
$k$ in this regime and yielding a white-noise scaling,
\begin{equation}
    \widetilde\Delta_X^2(k,\eta) \;\propto\; k^3
    \qquad(k\ll k_{\rm end})\,.
    \label{eq:k3_scaling}
\end{equation}
This $k^3$ behavior is the maximally blue-tilted limit of the spectral
index $n_\chi-1\to 3$ discussed in
Section~\ref{sec:spectator}, and arises naturally when the
portal coupling dominates the effective mass during inflation
($\sigma\phi_*^2\gg m_\chi^2,\,H_*^2$).  Since the evolution of all
IR modes is identical, we can factorize the time dependence by defining
a transfer function
\begin{equation}
    \widetilde\Delta_X^2(k,\eta)
    \;=\;
    \mathcal{T}_X(\eta)\,\widetilde\Delta_X^2(k)
    \qquad(k\ll k_{\rm end})\,,
    \label{eq:TX}
\end{equation}
where $\mathcal{T}_X(\eta)\to 1$ at late times by construction, and
$\widetilde\Delta_X^2(k)$ is the frozen (final) spectrum.  This
factorization is exact in the IR and breaks down only for modes with
$k\gtrsim k_{\rm end}$, where the $k^2$ gradient term in $\omega_k^2$
introduces mode-dependent evolution.

\begin{figure*}[t!]
    \centering
    \includegraphics[width=\linewidth]{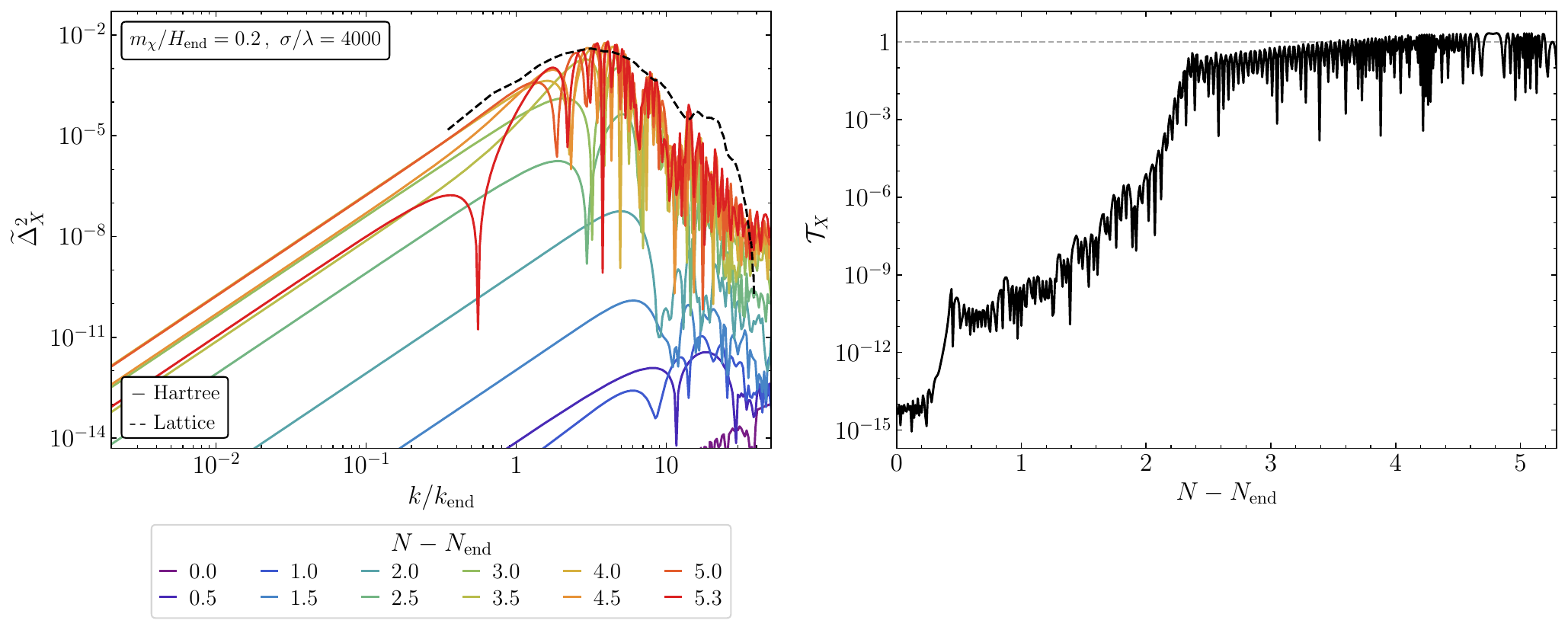}
    \caption{\textbf{Left:} Time evolution of the rescaled power
    spectrum $\widetilde\Delta_X^2(k,\eta)$ defined in
    Eq.~\eqref{eq:tildeDeltaX}, shown at successive epochs during
    reheating (colored solid curves, labeled by $e$-folds after the end
    of inflation $N-N_{\rm end}$).  The benchmark parameters are
    $m_\chi/H_{\rm end}=0.2$ and $\sigma/\lambda=4000$. The
    spectrum grows by roughly 15 orders of magnitude over
    $\sim 5$~$e$-folds, driven by parametric amplification from the
    oscillating inflaton portal mass $\sigma\phi^2(\eta)$. At low $k$
    the spectrum follows a white-noise scaling
    $\widetilde\Delta_X^2\propto k^3$, while a broad peak develops near
    $k\sim k_{\rm end}$. The dashed black curve shows the final
    spectrum obtained from a full nonlinear lattice simulation with
    \textsc{CosmoLattice}:
    the agreement with the Hartree result (solid curves) is excellent
    for $k\lesssim k_{\rm end}$, while additional UV power from
    mode-mode rescattering appears at
    $k\gtrsim\text{few}\times k_{\rm end}$.
    \textbf{Right:} Transfer function $\mathcal{T}_X(\eta)$ defined in
    Eq.~\eqref{eq:TX}, encoding the universal time evolution of all
    infrared modes ($k\ll k_{\rm end}$). The rapid oscillations track
    the inflaton oscillation frequency, with each half-period producing
    a burst of non-adiabatic amplification. The transfer function
    saturates at $\mathcal{T}_X\simeq 1$ (dashed gray line) once the
    inflaton amplitude has decayed sufficiently for the spectator
    evolution to become adiabatic.}
    \label{fig:PX_T}
\end{figure*}

Fig.~\ref{fig:PX_T} illustrates the evolution for the benchmark
$m_\chi/H_{\rm end}=0.2$, $\sigma/\lambda=4000$.\footnote{We note that this $\lambda$ arises from the inflaton potential~(\ref{eq:tmodel_potential}) and should not be confused with the self-interaction of a spectator scalar field $\lambda_{\chi}$.}  The left panel
shows $\widetilde\Delta_X^2(k,\eta)$ at several epochs during
reheating ($N-N_{\rm end}=0$-$5.3$).  At the end of inflation
($N-N_{\rm end}=0$), the spectrum is extremely suppressed: the large
portal coupling $\sigma\phi_*^2\gg H_*^2$ has kept the effective mass
heavy during inflation, strongly suppressing superhorizon growth. Over
the subsequent $\sim 5$~$e$-folds, the spectrum grows by
$\sim 15$~orders of magnitude. This dramatic amplification is driven
by the oscillating inflaton condensate: as $\phi(\eta)$ oscillates in
the quadratic minimum, the time-dependent portal mass
$\sigma\phi^2(\eta)$ periodically drives $\omega_k^2$ through
non-adiabatic regimes, parametrically amplifying the spectator mode
functions. This dramatic amplification is driven by the parametric resonance
mechanism described in Section~\ref{sec:gpp}: as $\phi(\eta)$
oscillates in the quadratic minimum, the time-dependent portal mass
$\sigma\phi^2(\eta)$ periodically drives $\omega_k^2$ through
non-adiabatic regimes, amplifying the spectator mode functions.

The dashed curve in the left panel shows the final
($N-N_{\rm end}=5.3$) spectrum obtained from a full nonlinear lattice
simulation using
\textsc{CosmoLattice}~\cite{Figueroa:2020rrl,Figueroa:2021yhd}.\footnote{The
  lattice simulation includes the complete nonlinear dynamics of both
  the inflaton and spectator, including rescattering, fragmentation,
  and backreaction effects beyond the Hartree approximation.}  The
agreement with the Hartree calculation is excellent for
$k\lesssim k_{\rm end}$, validating the mean-field treatment in the IR
where the bulk of the spectral weight resides.  At
$k\gtrsim\text{few}\times k_{\rm end}$, the lattice spectrum develops
additional UV power not captured by the Hartree approximation. 
This
arises from mode-mode rescattering and fragmentation that
redistributes power to shorter wavelengths. Since the scalar-induced
GW signal is dominated by modes near the spectral peak
($k\sim k_{\rm end}$, as we show in Section~\ref{sec:gwsignals}), the Hartree
treatment provides a reliable and computationally efficient
approximation for our purposes.

The right panel shows the corresponding transfer function
$\mathcal{T}_X(\eta)$. The oscillatory substructure tracks the
inflaton oscillation frequency, with each half-period of $\phi$
producing a burst of non-adiabatic amplification. The envelope grows
approximately exponentially before saturating at
$\mathcal{T}_X\simeq 1$ once the inflaton amplitude has decayed sufficiently for the evolution to become adiabatic.

%%%%%%%%%%%%%%%%%%%%%%%%%%%%%%%%%%%%%%%%%%%%%%%%%%%%%
\section{Scalar-Induced Gravitational Waves}
\label{sec:SIGWs}
%%%%%%%%%%%%%%%%%%%%%%%%%%%%%%%%%%%%%%%%%%%%%%%%%%%%%

%%%%%%%%%%%%%%%%%%%%%%%%%%%%%%%%%%%%%%%%%%%%%%%%%%%%%
\subsection{Gravitational Wave Spectral Density}
\label{sec:gwspectraldensity}
%%%%%%%%%%%%%%%%%%%%%%%%%%%%%%%%%%%%%%%%%%%%%%%%%%%%%
In this section we fix the conventions used to characterize the stochastic gravitational wave background (SGWB), following the standard treatments in Refs.~\cite{Maggiore:2007ulw,Domenech:2021ztg}. We parameterize scalar and tensor perturbations around a spatially flat FLRW background in conformal Newtonian gauge,
\begin{equation}
\label{eq:lineelement}
ds^2
\;=\;
a(\eta)^2\Big[
-(1+2\Phi)\,d\eta^2
+\big((1-2\Phi)\delta_{ij}+h_{ij}\big)\,dx^i dx^j
\Big],
\end{equation}
where $\Phi(\eta,\mathbf{x})$ is the Bardeen potential (we assume negligible anisotropic stress, so that $\Psi=\Phi$) and $h_{ij}(\eta,\mathbf{x})$ is the transverse-traceless (TT) tensor perturbation,
$\partial_i h_{ij}=0$, $h_i^{\ i}=0$.
We decompose the tensor field into Fourier modes and polarizations,
\begin{equation}
\label{eq:hijfour}
h_{ij}(\eta,\mathbf{x})
\;=\;
\sum_{\lambda=+,\times}\int\frac{d^3k}{(2\pi)^{3/2}}\,
e^{-i\mathbf{k}\cdot\mathbf{x}}\,
\varepsilon^{\lambda}_{ij}(\mathbf{k})\,h_\lambda(\eta,\mathbf{k})\,,
\end{equation}
where the polarization tensors satisfy $k^i\varepsilon_{ij}^\lambda=0$, $\varepsilon_{ii}^\lambda=0$, and $\varepsilon_{ij}^{\lambda}\varepsilon^{\lambda' ij}=\delta^{\lambda\lambda'}$, with $h_\lambda^*(\eta,\mathbf{k})=h_\lambda(\eta,-\mathbf{k})$ ensuring reality.  Statistical isotropy and homogeneity imply
\begin{equation}
\label{eq:tensor_2pt}
\langle h_\lambda(\eta,\mathbf{k})\,h_{\lambda'}(\eta,\mathbf{k}')\rangle
\;=\;
\delta_{\lambda\lambda'}\,\delta^{(3)}(\mathbf{k}+\mathbf{k}')\,\mathcal{P}_\lambda(\eta,k)\,,
\end{equation}
from which we define the dimensionless power spectrum per polarization and the total tensor power,
\begin{equation}
\Delta_\lambda^2(\eta,k)\;\equiv\;\frac{k^3}{2\pi^2}\,\mathcal{P}_\lambda(\eta,k)\,,
\qquad
\Delta_h^2(\eta,k)\;\equiv\;\sum_\lambda \Delta_\lambda^2(\eta,k)\,.
\label{eq:gwsgen}
\end{equation}

On sub-Hubble scales ($k\gg aH$), tensor perturbations propagate as free gravitational waves.  Averaging the quadratic GW energy-momentum tensor over several wavelengths (Isaacson prescription~\cite{Isaacson:1968hbi,Isaacson:1968zza}) and using the free-wave relation $|h_\lambda'|\simeq k|h_\lambda|$ gives the standard result for the GW energy density per logarithmic wavenumber (see e.g.\ Ref.~\cite{Maggiore:2007ulw} for the full derivation),
\begin{equation}
\frac{d\rho_{\rm GW}}{d\ln k}
\;\simeq\;
\frac{M_P^2\,k^2}{2 a^2}\,
\frac{k^3}{2\pi^2}\,\sum_\lambda \mathcal{P}_\lambda(\eta,k)\,.
\end{equation}
The dimensionless GW spectral density is therefore
\begin{equation}
\label{eq:gwgeneral}
\Omega_{\rm GW}(\eta,k)
\;\equiv\;
\frac{1}{\rho_{\rm tot}}\,\frac{d\rho_{\rm GW}}{d\ln k}
\;=\;
\frac{1}{12}\left(\frac{k}{a(\eta)H(\eta)}\right)^2
\Delta_h^2(\eta,k)\,,
\end{equation}
which is the form used throughout the paper.\footnote{This expression is exact for freely propagating modes well inside the Hubble radius.  During active GW production (e.g.\ while the scalar source is significant at horizon re-entry), the free-wave relation $|h_\lambda'|\simeq k|h_\lambda|$ does not hold and $\Omega_{\rm GW}$ should be evaluated from the full Isaacson expression.  In practice, we evaluate Eq.~\eqref{eq:gwgeneral} only after the source term has become negligible.}

Evaluated today, with $\rho_c = 3H_0^2M_P^2$, Eq.~\eqref{eq:gwgeneral} gives
\begin{equation}
\label{eq:gwpresentday}
\Omega_{{\rm GW},0} \; = \;
\frac{1}{12}\left(\frac{k}{a_0 H_0}\right)^2
\Delta_h^2(\eta_0,k)\,.
\end{equation}
The comoving wavenumber $k$ is related to the observed GW frequency by $f=k/(2\pi a_0)$, and the spectrum produced at early times is related to the present-day value by the standard transfer factor accounting for the change in $g_*$ and $g_{*s}$ between production and today~\cite{Watanabe:2006qe,Saikawa:2018rcs},
\begin{equation}
\label{eq:OmGW_transfer}
\Omega_{{\rm GW},0}\,h^2
\;=\;
\Omega_{r,0}\,h^2\;\times\;
\frac{g_{*}(T_{\rm prod})}{g_{*,0}}
\left(\frac{g_{*s,0}}{g_{*s}(T_{\rm prod})}\right)^{4/3}
\Omega_{\rm GW}(\eta_{\rm prod},k)\,,
\end{equation}
where $\Omega_{r,0}\,h^2\simeq 4.2\times 10^{-5}$, $g_{*,0}=3.36$, $g_{*s,0}=3.91$, and $T_{\rm prod}$ is the temperature at which the GW mode is produced (or, more precisely, enters the radiation-dominated regime).  For modes produced well above the electroweak scale, the ratio of $g_*$ factors gives an $\mathcal{O}(0.4)$ suppression relative to $\Omega_{r,0}$. 

In spectator field scenarios, $\Omega_{\rm GW}$ can encode both the inflationary initial conditions and the post-inflationary processing of scalar fluctuations (reheating dynamics, equation-of-state transitions, parametric amplification) that imprint characteristic spectral features.  In the next section we compute the scalar source that drives the production of $h_{ij}$ at second order.

%%%%%%%%%%%%%%%%%%%%%%%%%%%%%%%%%%%%%%%%%%%%%%%%%%%%%%%%%%%%%%%%%%
\subsection{Secondary Gravitational Waves from Scalar Perturbations}
\label{sec:gravwaves1}
%%%%%%%%%%%%%%%%%%%%%%%%%%%%%%%%%%%%%%%%%%%%%%%%%%%%%%%%%%%%%%%%%%
Scalar perturbations source tensor modes at second order in cosmological perturbation theory~\cite{Ananda:2006af,Baumann:2007zm,Domenech:2021ztg}. In this subsection we derive the equation of motion for the secondary GWs sourced by the spectator field fluctuations.  For the complementary signal arising from the evolving isocurvature power spectrum through $\Phi$-sourced induced GWs, typically at much lower frequencies, see Refs.~\cite{Ebadi:2023xhq,Garcia:2025yit,Garcia:2025wmu}.

Working in the Newtonian gauge~\eqref{eq:lineelement} with
$\Psi=\Phi$ (exact at first order in the absence of anisotropic
stress), the second-order Einstein equations yield an equation of the
form~\cite{Ananda:2006af,Baumann:2007zm,Domenech:2021ztg}
\begin{equation}
\label{eq:hij_eom_full}
    h_{ij}'' + 2\mathcal{H}\,h_{ij}' - \nabla^2 h_{ij}
    \;=\;
    \mathcal{S}_{ij}^{(\Phi)}
    \;+\;
    \frac{2}{M_P^2}\,
    \mathcal{P}^{ab}_{~~ij}\,
    \partial_a\delta\chi\,\partial_b\delta\chi\,,
\end{equation}
where all spatial derivatives are with respect to comoving coordinates
and $\mathcal{P}^{ab}_{~~ij}$ is the transverse-traceless projection
operator (defined below).  The first source,
$\mathcal{S}_{ij}^{(\Phi)}$, collects the terms quadratic in $\Phi$
and $\Phi'$ (with $w$-dependent coefficients) in which $\Phi$ is
sourced at first order by the inflaton perturbation $\delta\phi$. Its
explicit Fourier-space form is given in
Refs.~\cite{Ananda:2006af,Baumann:2007zm,Kohri:2018awv,Domenech:2021ztg}.
The second source, $\partial\delta\chi\,\partial\delta\chi$, arises
from the second-order stress-energy of the spectator
field.\footnote{Since the spectator is subdominant
  ($\rho_\chi\ll\rho_{\rm tot}$) and has vanishing background value
  ($\langle\chi\rangle=0$), it does not source $\Phi$ at linear
  order.  The $\delta\chi\delta\chi$ contribution to $h_{ij}$ is
  therefore independent of the metric-sourced terms, with no
  double-counting.  Metric-field cross terms such as
  $\Phi\,\partial^2(\delta\chi^2)$ are suppressed by
  $\rho_\chi/\rho_{\rm tot}\ll 1$ and are neglected.}

In this work we focus on the high-frequency SGWB generated by the
direct spectator source, which probes the small-scale spectator field power spectrum discussed in Section~\ref{sec:matterPS}. The complementary
metric-sourced contribution, which probes larger scales and was
analyzed in Refs.~\cite{Ebadi:2023xhq,Garcia:2025yit,Garcia:2025wmu}, is
neglected here.  The equation we solve is therefore
\begin{equation}
\label{eq:hij_eom_chi}
    h_{ij}'' + 2\mathcal{H}\,h_{ij}' - \nabla^2 h_{ij}
    \;=\;
    \frac{2}{M_P^2}\,\mathcal{P}^{ab}_{~~ij}\,
    \partial_a\chi\,\partial_b\chi\,,
\end{equation}
where we have set $\delta\chi=\chi$ since $\langle\chi\rangle=0$.

The TT projection operator in Fourier space is
\begin{equation}
\label{eq:TTprojector}
    \mathcal{P}^{ab}_{~~ij}(\hat{\mathbf{k}})
    \;=\;
    P_i^{\ a}(\hat{\mathbf{k}})\,P_j^{\ b}(\hat{\mathbf{k}})
    - \frac{1}{2}\,P_{ij}(\hat{\mathbf{k}})\,P^{ab}(\hat{\mathbf{k}})\,,
    \qquad
    P_{ij}(\hat{\mathbf{k}}) \;\equiv\; \delta_{ij} - \hat k_i\hat k_j\,.
\end{equation}
Projecting Eq.~\eqref{eq:hij_eom_chi} onto polarization $\lambda$
and passing to Fourier space as described above, we obtain
\begin{equation}
\label{eq:hijgensol}
    h_\lambda''(\eta,\mathbf{k})
    + 2\mathcal{H}\,h_\lambda'(\eta,\mathbf{k})
    + k^2\,h_\lambda(\eta,\mathbf{k})
    \;=\;
    S_\lambda^{(\chi)}(\eta,\mathbf{k})\,,
\end{equation}
where the source from the direct field-gradient channel is
\begin{equation}
\label{eq:source_chi}
    S_\lambda^{(\chi)}(\eta,\mathbf{k})
    \;=\;
    \frac{2}{a^2(\eta)\,M_P^2}
    \int\frac{d^3p}{(2\pi)^{3/2}}\,
    \varepsilon_\lambda^{ij}(\mathbf{k})\,p_i\,p_j\;
    \left[\hat X_{\mathbf{p}}(\eta)+\hat X_{-\mathbf{p}}^\dagger(\eta)\right]
    \left[\hat X_{\mathbf{k}-\mathbf{p}}(\eta)
    +\hat X_{-(\mathbf{k}-\mathbf{p})}^\dagger(\eta)\right].
\end{equation}
To arrive at Eq.~\eqref{eq:source_chi} we have used
$\delta\chi=\chi$ (since $\langle\chi\rangle=0$), inserted the mode
expansion $\chi=a^{-1}\int\frac{d^3p}{(2\pi)^{3/2}}
e^{-i\mathbf{p}\cdot\mathbf{x}}
[\hat X_\mathbf{p}+\hat X_{-\mathbf{p}}^\dagger]$,\footnote{For simplicity, we define $\hat X_{\mathbf{p}} = X_{p}(\eta) \hat{a}_{\bf p}$ and $\hat X^{\dagger}_\mathbf{-p} = X_{p}(\eta)^* \hat{a}^{\dagger}_{\bf -p}$ Compare with Eq.~(\ref{eq:xfourier})} and applied the
transversality identity
$\varepsilon_\lambda^{ij}(\mathbf{k})\,p_i(k_j-p_j)
=-\varepsilon_\lambda^{ij}(\mathbf{k})\,p_i p_j$
(since $\varepsilon_\lambda^{ij}k_j=0$) to eliminate the
$\mathbf{k}$-dependence from the integrand of the
convolution.\footnote{On sub-Hubble scales where the spectator
dominates the density perturbation, $\Phi$ is itself sourced by
$\delta\rho_\chi$ through the Poisson equation, giving rise to an
additional metric-channel contribution
$S_\lambda^{(\Phi)}\propto \delta\rho_\chi\,\delta\rho_\chi/(M_P^4 p^2|\mathbf{k}-\mathbf{p}|^2)$. This contribution, analyzed in Refs.~\cite{Ebadi:2023xhq,Garcia:2025yit,Garcia:2025wmu}, is neglected here. We focus exclusively on the direct field-gradient source~\eqref{eq:source_chi}, which dominates the high-frequency signal during the early stages when the spectator is still subdominant in the background.}

The solution to Eq.~\eqref{eq:hijgensol} with vanishing initial
conditions, $h_\lambda(\eta_i,\mathbf{k})
=h_\lambda'(\eta_i,\mathbf{k})=0$, is given by the retarded
Green's function integral
\begin{equation}
\label{eq:hk_Green}
    h_\lambda(\eta,\mathbf{k})
    \;=\;
    \int_{\eta_i}^{\eta}d\tilde\eta\;
    \mathcal{G}_{\mathbf{k}}(\eta,\tilde\eta)\,
    S_\lambda^{(\chi)}(\tilde\eta,\mathbf{k})\,,
\end{equation}
where $\mathcal{G}_{\mathbf{k}}(\eta,\tilde\eta)$ satisfies
\begin{equation}
\label{eq:Green_eq}
    \mathcal{G}_{\mathbf{k}}''(\eta,\tilde\eta)
    + 2\mathcal{H}\,\mathcal{G}_{\mathbf{k}}'(\eta,\tilde\eta)
    + k^2\,\mathcal{G}
    _k(\eta,\tilde\eta)
    \;=\;
    \delta(\eta-\tilde\eta)\,,
\end{equation}
with $\mathcal{G}_{\mathbf{k}}(\tilde\eta,\tilde\eta)=0$ and
$\mathcal{G}_{\mathbf{k}}'(\tilde\eta^+,\tilde\eta)=1$. Explicitly, if $h_{\lambda}^{(1)}(\eta)$ and
$h_{\lambda}^{(2)}(\eta)$ are two linearly independent homogeneous solutions of the $h_\lambda$, the Green's function reads
\begin{equation}
\label{eq:Green_explicit}
    \mathcal{G}_{\mathbf{k}}(\eta,\tilde\eta)
    \;=\;
    \frac{ h_{\lambda}^{(2)}(\eta)\,h_{\lambda}^{(1)}(\tilde\eta)-h_{\lambda}^{(1)}(\eta)\,h_{\lambda}^{(2)}(\tilde\eta)
          }
         {h_{\lambda}^{(1)}(\tilde\eta)\,h_{\lambda}^{(2)\,\prime}(\tilde\eta)
          - h_{\lambda}^{(1)\,\prime}(\tilde\eta)\,h_{\lambda}^{(2)}(\tilde\eta)}\,.
\end{equation}

%%%%%%%%%%%%%%%%%%%%%%%%%%%%%%%%%%%%%%%%%%%%%%%%%%%%%%%%%%%%%%%%%%
\subsection{Gravitational Wave Power Spectrum}
\label{sec:compt4point}
%%%%%%%%%%%%%%%%%%%%%%%%%%%%%%%%%%%%%%%%%%%%%%%%%%%%%%%%%%%%%%%%%%
We now compute the GW power spectrum by evaluating the four-point correlator of the spectator field that enters the two-point function of $h_\lambda$.

The angular coupling of the scalar source to GW polarization~$\lambda$ is encoded in the projection factor
\begin{equation}
\label{eq:Qlambda_def}
    Q_\lambda(\mathbf{k},\mathbf{q})
    \;\equiv\;
    \varepsilon_\lambda^{ij}(\mathbf{k})\,q_i\,q_j\,.
\end{equation}
Choosing $\hat{\mathbf{k}}=\hat z$ and writing
$\mathbf{q}=q(\sin\theta\cos\phi,\,\sin\theta\sin\phi,\,\cos\theta)$,
\begin{equation}
\label{eq:Qplus_Qcross}
    Q_+(\mathbf{k},\mathbf{q})
    = \frac{q^2\sin^2\!\theta}{\sqrt{2}}\,\cos 2\phi\,,
    \qquad
    Q_\times(\mathbf{k},\mathbf{q})
    = \frac{q^2\sin^2\!\theta}{\sqrt{2}}\,\sin 2\phi\,.
\end{equation}
Since $\varepsilon_\lambda^{ij}k_j=0$, the projection factor satisfies
$Q_\lambda(\mathbf{k},\mathbf{q})
=Q_\lambda(\mathbf{k},\mathbf{q}+c\mathbf{k})$ for any constant~$c$,
and is symmetric under $\mathbf{q}\to-\mathbf{q}$ and
$\mathbf{k}\to-\mathbf{k}$.  Summing over polarizations and averaging
over the azimuthal angle,
\begin{equation}
\label{eq:Qsum}
    \sum_\lambda\!\int_0^{2\pi}\!\frac{d\phi}{2\pi}\;
    Q_\lambda^2(\mathbf{k},\mathbf{q})
    \;=\;
    \frac{q^4\sin^4\!\theta}{2}\,.
\end{equation}

The GW power spectrum~\eqref{eq:gwsgen} requires the two-point
function of $h_\lambda$. Inserting the Green's function
solution~\eqref{eq:hk_Green} with the
source~\eqref{eq:source_chi},
\begin{equation}
\label{eq:hh_Green}
\begin{aligned}
    &\langle h_\lambda(\eta,\mathbf{k})\,
    h_\lambda(\eta,\mathbf{k}')\rangle
    \;=\;\;
    \frac{4}{M_P^4}
    \int_{\eta_i}^{\eta}\!d\eta_1
    \int_{\eta_i}^{\eta}\!d\eta_2\;
    \mathcal{G}_{\mathbf{k}}(\eta,\eta_1)\,\mathcal{G}_{\mathbf{k}'}(\eta,\eta_2) \\
    &\times
    \int\frac{d^3q_1}{(2\pi)^{3/2}}
    \frac{d^3q_2}{(2\pi)^{3/2}}\;
    Q_\lambda(\mathbf{k},\mathbf{q}_1)\,
    Q_\lambda(\mathbf{k}',\mathbf{q}_2) \times
    \langle
    \chi_{\mathbf{q}_1}(\eta_1)\,
    \chi_{\mathbf{k}-\mathbf{q}_1}(\eta_1)\,
    \chi_{\mathbf{q}_2}(\eta_2)\,
    \chi_{\mathbf{k}'-\mathbf{q}_2}(\eta_2)
    \rangle\,,
\end{aligned}
\end{equation}
where $\chi_\mathbf{q}(\eta)
=[\hat X_\mathbf{q}(\eta)
+\hat X_{-\mathbf{q}}^\dagger(\eta)]/a(\eta)$ is the Fourier-space
field operator.  For Gaussian $\chi$ (with $\langle\chi\rangle=0$),
the four-point function decomposes into disconnected and connected
parts,
$\langle\chi\chi\chi\chi\rangle
=\langle\chi\chi\chi\chi\rangle_{\rm d}
+\langle\chi\chi\chi\chi\rangle_{\rm c}$.
The connected part is proportional to the trispectrum of $\chi$,
which provides a subleading correction to the GW signal for
perturbative
non-Gaussianities~\cite{Garcia-Bellido:2017aan,Adshead:2021hnm, Atal:2021jyo, Garcia-Saenz:2022tzu}. We focus on the disconnected (Gaussian) contribution, so that our GW
predictions represent conservative lower bounds.

The disconnected four-point function consists of three Wick pairings,
constructed from the free-field two-point function
$\langle\chi_\mathbf{q}(\eta_1)\,\chi_\mathbf{p}(\eta_2)\rangle
=\delta^{(3)}(\mathbf{q}+\mathbf{p})\,
X_q(\eta_1)\,X_q^*(\eta_2)/[a(\eta_1)\,a(\eta_2)]$:

\noindent\emph{Pairing~(i):}
$\langle\chi_{\mathbf{q}_1}\chi_{\mathbf{k}-\mathbf{q}_1}\rangle
\langle\chi_{\mathbf{q}_2}\chi_{\mathbf{k}'-\mathbf{q}_2}\rangle
\propto\delta^{(3)}(\mathbf{k})\,\delta^{(3)}(\mathbf{k}')$.
This vanishes at finite momentum and does not contribute to the power
spectrum.

\noindent\emph{Pairing~(ii):}
$\langle\chi_{\mathbf{q}_1}\chi_{\mathbf{q}_2}\rangle
\langle\chi_{\mathbf{k}-\mathbf{q}_1}
\chi_{\mathbf{k}'-\mathbf{q}_2}\rangle
\propto\delta^{(3)}(\mathbf{k}+\mathbf{k}')\,
\delta^{(3)}(\mathbf{q}_1+\mathbf{q}_2)$.
Integrating over $\mathbf{q}_2$ and using
$Q_\lambda(-\mathbf{k},-\mathbf{q})
=Q_\lambda(\mathbf{k},\mathbf{q})$ gives an integrand proportional
to $Q_\lambda^2(\mathbf{k},\mathbf{q})\,
X_q(\eta_1)X_q^*(\eta_2)\,
X_{|\mathbf{k}-\mathbf{q}|}(\eta_1)
X_{|\mathbf{k}-\mathbf{q}|}^*(\eta_2)$.

\noindent\emph{Pairing~(iii):}
$\langle\chi_{\mathbf{q}_1}\chi_{\mathbf{k}'-\mathbf{q}_2}\rangle
\langle\chi_{\mathbf{k}-\mathbf{q}_1}\chi_{\mathbf{q}_2}\rangle
\propto\delta^{(3)}(\mathbf{k}+\mathbf{k}')\,
\delta^{(3)}(\mathbf{q}_1-\mathbf{k}+\mathbf{q}_2)$.
After integrating over $\mathbf{q}_2$, the shift property
$Q_\lambda(\mathbf{k},\mathbf{q}-\mathbf{k})
=Q_\lambda(\mathbf{k},\mathbf{q})$ reduces this to an integrand
identical to pairing~(ii).

The source $\partial_a\chi\,\partial_b\chi$ is a composite operator
whose vacuum expectation value diverges in the ultraviolet.
Physically, the zero-point fluctuations of $\chi$ do not source
gravitational waves; the vacuum stress-energy is removed by the
standard cosmological-constant renormalization.  This is implemented
by normal-ordering the source, which replaces
$|X_q|^2\to|X_q|^2-1/(2\omega_q)$ in each leg of the Wick
contraction, precisely the adiabatic vacuum subtraction used in
the definition of
$\Delta_X^2$~\eqref{eq:DeltaX}.\footnote{The renormalization can
  equivalently be understood as the subtraction of the instantaneous
  WKB mode
  $f_q(\eta)=[2\omega_q(\eta)]^{-1/2}\exp(-i\!\int^\eta\!\omega_q)$
  in each Wick contraction leg, which at equal times reduces to
  $|X_q|^2-1/(2\omega_q)$.  This is the same adiabatic
  regularization used in defining the renormalized energy
  density~\cite{Birrell:1982ix,Parker:1974qw}.}  Without this
subtraction, the momentum integral in the GW power spectrum is
ultraviolet divergent: in the UV, $|X_q|^2\to 1/(2\omega_q)\sim
1/(2q)$ and the integrand grows as~$q^4$.

Combining the two equivalent pairings with the vacuum subtraction,
the renormalized source-source correlator becomes
\begin{equation}
\label{eq:SS_corr}
\begin{aligned}
    \langle S_\lambda^{(\chi)}(\eta_1,\mathbf{k})\,
    &S_\lambda^{(\chi)}(\eta_2,\mathbf{k}')\rangle
    \;=\;
    \frac{8\,\delta^{(3)}(\mathbf{k}+\mathbf{k}')}
         {a^2(\eta_1)\,a^2(\eta_2)\,M_P^4} \\
    &\times\int\!\frac{d^3q}{(2\pi)^3}\,
    Q_\lambda^2(\mathbf{k},\mathbf{q})\;
    \left(|X_q(\eta_1)|^2
    -\frac{1}{2\omega_q(\eta_1)}\right)
    \left(|X_{|\mathbf{k}-\mathbf{q}|}(\eta_2)|^2
    -\frac{1}{2\omega_{|\mathbf{k}-\mathbf{q}|}(\eta_2)}\right)  \\
     & = \frac{4\pi\,\delta^{(3)}(\mathbf{k}+\mathbf{k}')}
         {a^2(\eta_1)\,a^2(\eta_2)} \times\int\! {d^3q}\,
    \frac{Q_\lambda^2(\mathbf{k},\mathbf{q})}{q^3
 |\mathbf{k}- \mathbf{q}|^3}\;
   \Delta_X^2(q, \eta_1) \Delta_X^2(|\mathbf{k} - \mathbf{q}|, \eta_2) \, ,
\end{aligned}
\end{equation}
where the factor of~8 arises as $4\times 2$: four from
$(2/M_P^2)^2$ in the source~\eqref{eq:source_chi}, and two from the
equivalent Wick pairings. The unequal-time products have been
evaluated at equal times within each Wick pair,
$X_q(\eta_1)X_q^*(\eta_2)\to|X_q(\eta_1)|^2$ and similarly for
$\eta_2$, which is valid in the regime where the transfer function
decomposition~\eqref{eq:TX} holds and the time and momentum integrals
factorize.

Inserting Eq.~\eqref{eq:SS_corr} into Eq.~\eqref{eq:hh_Green},
extracting the dimensionless power spectrum via Eq.~\eqref{eq:gwsgen}, summing over both polarizations, and using the transfer function factorization to separate the frozen spectral content from the time-dependent build-up, we obtain
\begin{equation}
\label{eq:Deltah_master}
    \Delta_h^2(\eta, k)
    \;=\; \frac{2 k^3}{\pi}
    \int\! d^3q\;
    \frac{Q^2(\mathbf{k},\mathbf{q})}{
 q^3 |\mathbf{k}- \mathbf{q}|^3}\;
    \widetilde\Delta_X^2(q)\,
    \widetilde\Delta_X^2(|\mathbf{k} - \mathbf{q}|)
    \left|
    \int_{\eta_i}^{\eta}\!d\eta'\,
    \frac{\mathcal{G}_{\mathbf{k}}(\eta,\eta')\,
    \mathcal{T}_X(\eta')}{a^3(\eta')/a_{\rm end}^3}
    \right|^2,
\end{equation}
where $Q^2\equiv\sum_\lambda Q_\lambda^2=q^4\sin^4\!\theta/2$ after the
azimuthal average~\eqref{eq:Qsum}, and the vacuum-subtracted mode
functions are evaluated at late times (after freeze-out). The vacuum
subtraction ensures ultraviolet convergence of the momentum integral,
with only the produced quanta ($n_q\neq 0$) contributing to the GW
signal.

Performing the integration, we can write the above equations as:
\begin{equation}
\label{eq:Deltah_efold}
    \Delta_h^2(k,N)
    \;=\;
   \frac{1}{4}
    \left[ \int_0^{N}\!dN'\;
    \frac{\mathcal{G}_{\mathbf{k}}(N,N')\,\mathcal{T}_X(N')}
         {a(N')^3/a_{\rm end}^3}
    \left(\frac{k}{a(N')H(N')}\right)^{\!2}
    \right]^{\!2}
    g(k)\,,
\end{equation}
where we have switched to $e$-folds $N=\ln a$ as the time variable
(with $dN=Hdt$), $\mathcal{G}_{\mathbf{k}}(N,N')$ is the Green's function in
$e$-fold variables. The spectral integral is defined as
\begin{equation}
\label{eq:gkint}
    g(k)
    \;\equiv\;
    \frac{1}{k^6}
    \int_0^\infty\!dp
    \int_{|k-p|}^{k+p}\!dq\;
    \frac{\left[k^4-2k^2(p^2+q^2)
          +(p^2-q^2)^2\right]^2}
         {p^2\,q^2}\;
    \widetilde\Delta_X^2(q)\,
    \widetilde\Delta_X^2(p)\,.
\end{equation}

In the strong-coupling regime, the frozen spectrum
$\widetilde\Delta_X^2(k)\propto k^3$ for $k\ll k_{\rm peak}$ (with
the peak near $k_{\rm end}$, cf.\
Section~\ref{sec:matterPS}).  In this limit the spectral integral can
be evaluated analytically, giving
\begin{equation}
\label{eq:gk_analytic}
    g(k)
    \;\simeq\;
    \frac{256}{105}
    \left(\frac{k_{\rm peak}}{k_{\rm end}}\right)^{\!7}
    \left(\frac{k_{\rm end}}{k}\right)
    \left(\frac{\widetilde\Delta_X^2(k)}{k^3}\right)^{\!2}\,,
    \qquad(k\ll k_{\rm peak})\,.
\end{equation}
The $1/k$ scaling reflects the fact that for a white-noise input
$\widetilde\Delta_X^2\propto k^3$, the convolution
integral~\eqref{eq:gkint} is dominated by modes near the spectral
peak, whose contributions grow with the volume of the available
phase space as $k$ decreases.

\begin{figure}[!t]
\centering
    \includegraphics[width=0.8\textwidth]{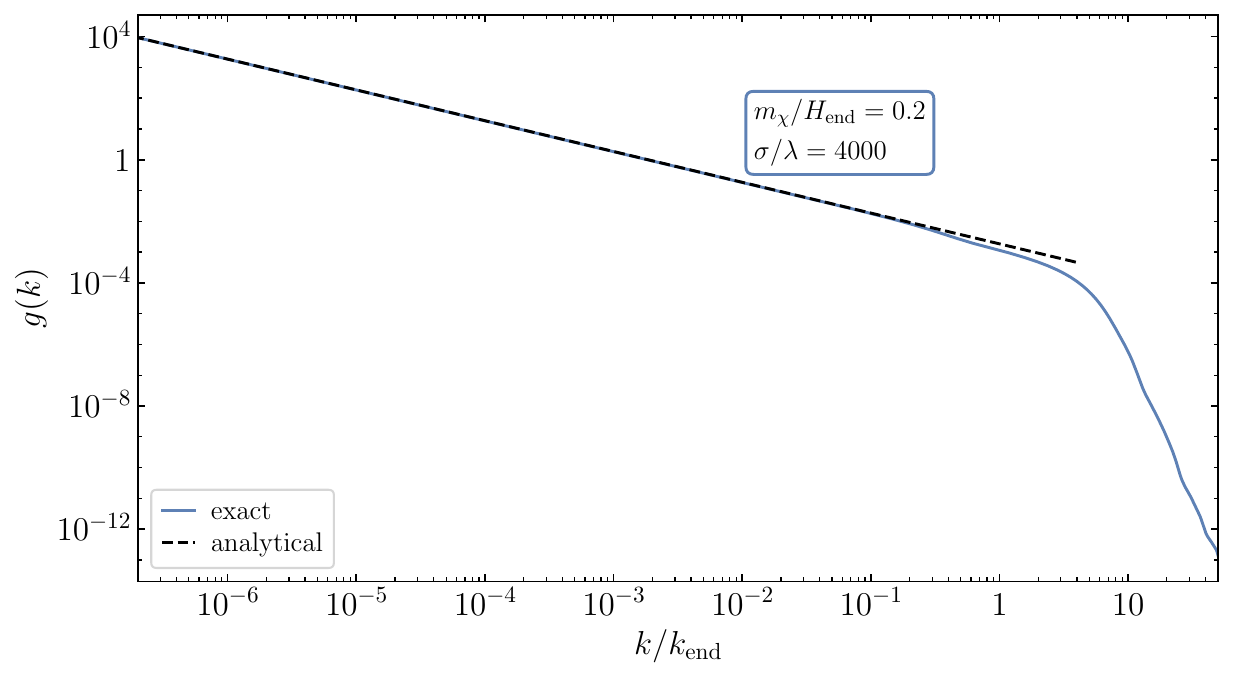}
    \caption{The spectral integral $g(k)$ defined in
    Eq.~\eqref{eq:gkint} for the benchmark parameters
    $m_\chi/H_{\rm end}=0.2$ and $\sigma/\lambda=4000$.  The
    solid blue curve shows the exact numerical evaluation; the
    dashed black line shows the analytical
    approximation~\eqref{eq:gk_analytic}, valid in the infrared
    regime $k\ll k_{\rm peak}\sim k_{\rm end}$ where
    $\widetilde\Delta_X^2\propto k^3$. The $g(k)\propto k^{-1}$
    scaling in the IR reflects the phase-space volume available in
    the convolution.  For $k\gtrsim k_{\rm end}$, the integrand
    probes the exponential falloff of $\widetilde\Delta_X^2$ beyond
    the spectral peak, leading to the sharp suppression and
    oscillatory features.}
    \label{fig:gk}
\end{figure}

Fig.~\ref{fig:gk} compares the exact numerical evaluation of
$g(k)$ with the analytical approximation~\eqref{eq:gk_analytic} for the benchmark $m_\chi/H_{\rm end}=0.2$, $\sigma/\lambda = 4000$.  The agreement is excellent over the entire infrared range $k\ll k_{\rm peak}\sim k_{\rm end}$, confirming the $k^{-1}$ scaling. For $k\gtrsim k_{\rm end}$, the spectrum probes the detailed shape of $\widetilde\Delta_X^2$ beyond the white-noise regime and $g(k)$ drops
sharply, with oscillatory features reflecting the structure of the spectral peak.

The master formula~\eqref{eq:Deltah_efold}, together with the
spectral integral~\eqref{eq:gkint} and the Green's functions evaluated in
Section~\ref{sec:gwmat}, provide the complete framework for computing
the scalar-induced GW signal from the spectator sector.  We evaluate
it numerically in Section~\ref{sec:gwsignals}.

%%%%%%%%%%%%%%%%%%%%%%%%%%%%%%%%%%%%%%%%%%%%%%%%%%%%%
\section{Gravitational Wave Production from Matter Perturbations}
\label{sec:gwmat}
%%%%%%%%%%%%%%%%%%%%%%%%%%%%%%%%%%%%%%%%%%%%%%%%%%%%%
The master formula~\eqref{eq:Deltah_efold} factorizes the GW power spectrum into a momentum integral $g(k)$~\eqref{eq:gkint} that encodes the frozen spectral shape of the spectator, and a time integral that encodes the build-up of the GW signal through the post-inflationary expansion history.  In this section we evaluate the time-dependent factor, exploiting the distinct scaling of the cosmological background in the matter-dominated (reheating) and radiation-dominated eras to obtain a semi-analytical estimate.

Using Eq.~\eqref{eq:Deltah_efold} together with the relation~\eqref{eq:gwpresentday}, and following the computation of the isocurvature-sourced spectrum in Ref.~\cite{Garcia:2025wmu}, we define the GW density parameter sourced by the spectator as
\begin{align}
\label{eq:OGWfact}
\Omega_{{\rm GW},X} \;&=\;  \frac{1}{12}\left(\frac{k}{aH}\right)^2\Delta_h^2(k, N)\;\equiv\;\mathcal{I}(k,N) g(k)\,.
\end{align}
with $g(k)$ defined in (\ref{eq:gkint}), and 
\begin{equation}
\mathcal{I}(k,N)
    \;\equiv\;
    \frac{1}{48} \left(\frac{k}{a(N)H(N)}\right)^2
    \left[ \int_0^{N}\!dN'\;
    \frac{\mathcal{G}_{\mathbf{k}}(N,N')\,\mathcal{T}_X(N')}
         {(a(N')/a_{\rm end})^3}
    \left(\frac{k}{a(N')H(N')}\right)^{\!2}
    \right]^{\!2}\,.
\end{equation}

A simplified form for this expression can be obtained by focusing on the distinct time-evolution of the cosmological background and the spectator field following the end of inflation.  Following our
discussion in Section~\ref{sec:gpp}, we assume that the adiabatic, non-relativistic regime for $\chi$ is reached before the end of reheating, occurring at the $e$-fold number $N_{\rm NR}$.  Prior to
this moment, the form of $\mathcal{T}_X$ cannot be determined analytically and requires the numerical computation of the spectator field power spectrum (see Fig.~\ref{fig:PX_T}). The GW Green's function and the background functions can also be fully extracted from the numerical solution of the dynamics, although we find that approximating them by their matter-domination forms provides an
accurate result. For $N_{\rm NR}<N<N_{\rm reh}$, we can continue to use the matter-domination approximation, as the inflaton continues to oscillate in its quadratic potential, with the added simplification that we can set $\mathcal{T}_X\simeq 1$.  After reheating, and up to the decay time of $\chi$ (if the spectator is unstable) or the onset of matter-radiation equality (if $\chi$ is stable), we can apply analogous simplifications, this time in the radiation-dominated background.  Specifically, we split $\mathcal{I}(k,N)$ into three parts
\begin{align} \notag
\mathcal{I}(k,N) \;&\simeq\; \frac{1}{48} \left(\frac{k}{a(N)H(N)}\right)^2\Bigg[\int_0^{N_{\rm NR}}dN'\,\frac{\mathcal{G}_{\bk}(N,N') \mathcal{T}_X(N') }{(a(N')/a_{\rm end})^3} \left( \frac{k}{a(N') H(N')} \right)^2 \\ \notag
&\hspace{130pt}+ \int_{N_{\rm NR}}^{N_{\rm reh}}dN'\,\frac{\mathcal{G}_{\bk}(N,N') }{(a(N')/a_{\rm end})^3} \left( \frac{k}{a(N') H(N')} \right)^2\\ \label{eq:I2x}
&\hspace{130pt} + \int_{N_{\rm reh}}^{N}dN'\, \frac{\mathcal{G}_{\bk}(N,N') }{(a(N')/a_{\rm end})^3} \left( \frac{k}{a(N') H(N')} \right)^2  \Bigg]^2\\ \label{eq:I2a}
&\equiv\; \bigg[ \mathcal{K}_{\rm NR}(k,N) + \mathcal{K}_{\rm reh}(k,N) + \mathcal{K}_{\rm rad}(k,N)\bigg]^2\,.
\end{align}
We now focus on the separate evaluation of each of the $\mathcal{K}$
contributions above.

%%%%%%%%%%%%%%%%%%%%%%%%%%%%%%%%%%%%%%%%%%%%%%%%%%%%%
\subsection{Early-time contribution ($\mathcal{K}_{\rm NR}(k,N)$)}
\label{sec:KNR}
%%%%%%%%%%%%%%%%%%%%%%%%%%%%%%%%%%%%%%%%%%%%%%%%%%%%%
We start with the early-time contribution $\mathcal{K}_{\rm NR}(k,N)$, obtained by evaluating the $N'$
integral in Eq.~\eqref{eq:I2x} up to the onset of the non-relativistic regime at $N_{\rm NR}$.  Including the appropriate prefactors, this corresponds to
\begin{align}\notag
\mathcal{K}_{\rm NR}(k,N) \;&=\;  \frac{1}{4\sqrt{3}} \left(\frac{k}{a(N)H(N)}\right) \int_0^{N_{\rm NR}}dN'\,\frac{\mathcal{G}_{\bk}(N,N') \mathcal{T}_{X}(N') }{(a(N')/a_{\rm end})^3} \left( \frac{k}{a(N') H(N')} \right)^2 \\ 
\label{eq:KNR1}
&\simeq\;  \frac{1 }{4\sqrt{3}} \left(\frac{k}{k_{\rm end}}\right)^3\left(\frac{a_{\rm end}H_{\rm end}}{a(N)H(N)}\right) \int_{N_{\rm end}}^{N_{\rm NR}}dN'\, \frac{\mathcal{G}_{\bk}(N,N') \mathcal{T}_{X}(N') }{(a(N')/a_{\rm end})^3} \left(\frac{a_{\rm end}H_{\rm end}}{a(N')H(N')}\right)^{2}\,,
\end{align}
where in the second line we have neglected the contribution from inflation, when the growth of spectator fluctuations is suppressed in the presence of large effective masses $\sigma\phi_*^2\gg H_*^2$. 

The Green's function for the tensor mode during the matter-like period of inflaton oscillations can be obtained analytically. Switching from conformal time to the $e$-fold variable in Eq.~\eqref{eq:hij_eom_chi} and expanding in Fourier modes, the homogeneous equation satisfied by the tensor modes is
\beq
\label{eq:hkode}
\frac{d^2 h_{\lambda}}{dN^2} + (3-\varepsilon_H)\frac{dh_{\lambda}}{dN} + \left(\frac{k}{aH}\right)^2 h_{\lambda} \;=\; 0\,,
\eeq
with $\varepsilon_H$ the first Hubble flow function,
\beq
\varepsilon_H \; = \; -\frac{\dot{H}}{H^2} \; = \; -\frac{1}{H}\frac{dH}{dN} \, .
\eeq
During the matter-dominated reheating epoch (\ref{eq:hkode}) simplifies to 
\begin{align} \label{eq:heqM}
\frac{d^2 h_{\lambda}}{dN^2} + \frac{3}{2}\frac{dh_{\lambda}}{dN} +\left(\frac{k}{k_{\rm end}}\right)^2 e^{(N-N_{\rm end})} h_{\lambda} \;=\;0\,,
\end{align}
with linearly independent solutions
\begin{equation}\label{eq:Gmatter}
    \begin{aligned}
        &h_{\lambda}^{(1)}(N,\bk) \;=\; e^{-\frac{3}{2}(N-N_{\rm end})}\left[ \cos\left(\frac{2k}{a(N)H(N)}\right) \right.  \left.+\; \frac{2k}{a(N)H(N)}\, \sin\left( \frac{2k}{a(N)H(N)} \right) \right]\,,\\ 
    &h_{\lambda}^{(2)}(N,\bk) \;=\; e^{-\frac{3}{2}(N-N_{\rm end})}\left[ \sin\left(\frac{2k}{a(N)H(N)}\right) \right. 
    \left.-\; \frac{2k}{a(N)H(N)}\, \cos\left( \frac{2k}{a(N)H(N)} \right) \right] \,,
    \end{aligned}
\end{equation}
where $a(N)H(N)=e^{-\frac{1}{2}(N-N_{\rm end})}a_{\rm end}H_{\rm end}$. The Green's function is then obtained by substitution into  (\ref{eq:Green_explicit}).

The prefactor and the first argument of the Green's function in Eq.~\eqref{eq:KNR1} are to be evaluated at the end of radiation domination, long after the modes have entered the horizon. To extend the Green's
function past $N_{\rm NR}$ and through the subsequent eras, we introduce the matter-domination and radiation-domination tensor transfer functions $\mathcal{T}_{\rm M}$ and $\mathcal{T}_{\rm R}$, respectively. Equation~\eqref{eq:KNR1} can then be rewritten as
\begin{align}\notag
\mathcal{K}_{\rm NR}(k,N) \;&\simeq\;  \frac{1}{4\sqrt{3}} \left(\frac{k}{k_{\rm end}}\right)^3 \left(\frac{a_{\rm end}H_{\rm end}}{a(N)H(N)}\right)  \mathcal{T}_{\rm R}(N;N_{\rm reh}) \mathcal{T}_{\rm M}(N_{\rm reh};N_{\rm NR})\\ \notag
&\hspace{70pt} \times \int_{N_{\rm end}}^{N_{\rm NR}}dN'\, \frac{\mathcal{G}_{\bk}(N_{\rm NR},N') \mathcal{T}_{X}(N') }{(a(N')/a_{\rm end})^3} \left(\frac{a_{\rm end}H_{\rm end}}{a(N')H(N')}\right)^{2}\,.
\end{align}

The matter-domination transfer function is obtained by integrating Eq.~\eqref{eq:heqM} for a solution of the form $h_\lambda(N,k)=\mathcal{T}_{\rm M}(N;N_{\rm NR})\, h_\lambda(N_{\rm NR},k)$, giving
\begin{align} \notag
&\mathcal{T}_{\rm M}(N;N_{\rm NR})\;=\; \left[ \frac{3}{8} \left(\frac{aH}{k}\right)^3 -\frac{1}{2}\left(\frac{aH}{ k}\right)\left(\frac{aH}{a_{\rm NR}H_{\rm NR}}\right)^2+\frac{3}{2}\left(\frac{aH}{ k}\right) \left(\frac{aH}{a_{\rm NR}H_{\rm NR}}\right) \right] \sin \left(2 \left(\frac{k}{aH}-\frac{k}{a_{\rm NR}H_{\rm NR}}\right)\right)\\  \displaybreak[0]
& +\left[ \left(\frac{aH}{a_{\rm NR}H_{\rm NR}}\right)^2 -\frac{3}{4} \left(\frac{aH}{k}\right)^2 + \frac{3}{4} \left(\frac{aH}{k}\right)^2 \left(\frac{aH}{a_{\rm NR}H_{\rm NR}}\right) \right] \cos \left(2 \left(\frac{k}{aH}-\frac{k}{a_{\rm NR}H_{\rm NR}}\right)\right)\\  \label{eq:TMapp}
&\simeq\; \begin{cases}
1\,, & \dfrac{k}{aH}\ll 1\,,\\[7pt]
 \left(\dfrac{aH}{a_{\rm NR}H_{\rm NR}}\right)^2\cos \left(2 \left(\dfrac{k}{aH}-\dfrac{k}{a_{\rm NR}H_{\rm NR}}\right)\right)\,, & \dfrac{k}{aH}\gg 1\,.
\end{cases}
\end{align}
In turn, to compute the radiation-domination transfer function $\mathcal{T}_{\rm R}$, we solve (\ref{eq:hkode}) in a radiation background,
\beq\label{eq:heqrad}
\frac{d^2 h_{\lambda}}{dN^2} + \frac{dh_{\lambda}}{dN} + \left(\frac{k}{k_{\rm reh}}\right)^2 e^{2(N-N_{\rm reh})} h_{\lambda} \;=\;0\,,
\eeq
where $k_{\rm reh}=a_{\rm reh}H_{\rm reh}$. Direct integration yields
\begin{align} 
\mathcal{T}_{\rm R}(N;N_{\rm reh}) \;=\; \left(\frac{aH}{a_{\rm reh}H_{\rm reh}}\right) \left[ \cos\left(\frac{k}{k_{\rm reh}}\left(1 - \frac{a_{\rm reh}H_{\rm reh}}{aH}\right)\right) \right. \left. - \left(\frac{k_{\rm reh}}{k}\right)
\sin\left(\frac{k}{k_{\rm reh}}\left(1 - \frac{a_{\rm reh}H_{\rm reh}}{aH}\right)\right)
\right] \, .
\label{eq:TRcompact}
\end{align}
At the present epoch all observationally relevant modes are well inside the horizon, with $k\gg a_0H_0$. Applying this approximation the transfer function simplifies to
\begin{align} \mathcal{T}_{\rm R}(N;N_{\rm reh}) \;\simeq\; \left(\frac{aH}{a_{\rm reh}H_{\rm reh}}\right) \cos \left(\frac{k}{k_{\rm reh}} \left(1-\frac{a_{\rm reh}H_{\rm reh}}{aH}\right)\right)\,. 
\end{align}
Upon averaging over the fast oscillations we obtain
\beq\label{eq:TRav}
\overline{\mathcal{T}_{\rm R}(N;N_{\rm reh})} \;\simeq\; \frac{1}{\sqrt{2}}\left(\frac{aH}{a_{\rm reh}H_{\rm reh}}\right)\,.
\eeq
With this, we can finally simplify the early-time contribution to $\Omega_{{\rm GW},X}$ as
\begin{align}\label{eq:KNRf}
\mathcal{K}_{\rm NR}(k,N) \;&\simeq\;  \frac{1}{4\sqrt{6}} \left(\frac{k}{k_{\rm end}}\right)^3 \left(\frac{a_{\rm end}H_{\rm end}}{a_{\rm reh}H_{\rm reh}}\right) \mathcal{T}_{\rm M}(N_{\rm reh};N_{\rm NR})\\ \notag
&\hspace{70pt} \times \int_{N_{\rm end}}^{N_{\rm NR}}dN'\,\frac{\mathcal{G}_{\bk}(N_{\rm NR},N') \mathcal{T}_{X}(N') }{(a(N')/a_{\rm end})^3} \left(\frac{a_{\rm end}H_{\rm end}}{a(N')H(N')}\right)^{2}\,.
\end{align}

\begin{figure}[!t]
\centering
    \includegraphics[width=\textwidth]{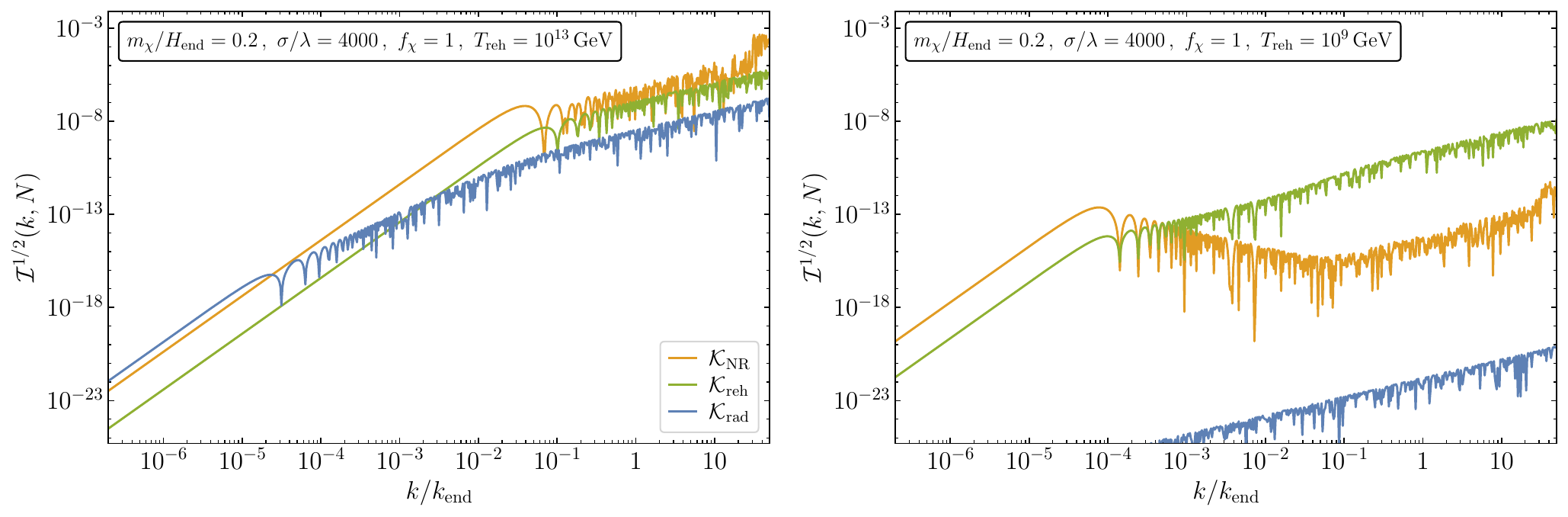}
    \caption{Contributions to the time-dependent factor
    $\mathcal{I}(k,N)$ of the GW energy spectrum, evaluated at the
    present epoch for two representative reheating temperatures: $T_{\rm reh}=10^{13}\,{\rm GeV}$ (left) and $T_{\rm reh}=10^{9}\,{\rm GeV}$ (right).
    The spectator parameters are $m_\chi/H_{\rm end}=0.2$ and
    $\sigma/\lambda=4000$, and the spectator is assumed to
    decay into radiation at $\rho_\chi=\rho_r$ (curvaton-like
    scenario with $f_\chi\equiv\rho_\chi/\rho_r=1$ at decay).  The
    plot shows the relative magnitudes of the early-time
    contribution ($\mathcal{K}_{\rm NR}$, orange), the reheating
    contribution ($\mathcal{K}_{\rm reh}$, green), and the radiation
    contribution ($\mathcal{K}_{\rm rad}$, blue).}
    \label{fig:ks}
\end{figure}

The result of integrating (\ref{eq:KNRf}) for the two reheating temperatures $T_{\rm reh}=10^{13},10^9\,{\rm GeV}$ is shown in Fig.~\ref{fig:ks} as the orange curves. Here we have chosen for definiteness a spectator with $m_{\chi}=0.2H_{\rm end}$ and $\sigma/\lambda=4000$. The dependence on these parameters is weak, being relevant for the determination of the value of $N_{\rm NR}$. Also for definiteness we have chosen a curvaton-like scenario for which the decay into radiation occurs at the latest possible moment, with $f_{\chi}=1$. We note that the maximum amplitude of $\mathcal{K}_{\rm NR}$ follows a direct relation to $T_{\rm reh}$, suggesting that the GW signal is maximal for the highest possible reheating temperatures. Similarly, we note a smooth behavior for modes that are outside the horizon during reheating $k\ll k_{\rm reh}$. In this regime, the Green's function has the simple scale-invariant form
\beq\label{eq:Gkrehapp}
\mathcal{G}_{\bk}(N,N')\;\simeq\; \frac{2}{3}\left( 1- e^{\frac{2}{3}(N'-N)} \right)\,.\qquad (k\ll k_{\rm reh})
\eeq
With $\mathcal{T}_{\rm M}(N_{\rm reh};N_{\rm NR})\simeq 1$ for these scales, we conclude that $\mathcal{K}_{\rm NR}(k,N)\propto k^3$, which is confirmed by the slope of the curves shown in the Figure. The amplitude requires numerical evaluation due to the presence of $\mathcal{T}_X(N')$ in the integrand of (\ref{eq:KNRf}). Nevertheless, we note that this amplitude is proportional to that of $\mathcal{K}_{\rm reh}$, which we will evaluate analytically in the next subsection. For smaller scales ($k\gg k_{\rm reh}$) the form of $\mathcal{K}_{\rm NR}$ is complicated due to the momentum dependence of the Green's and transfer functions in (\ref{eq:KNRf}).

%%%%%%%%%%%%%%%%%%%%%%%%%%%%%%%%%%%%%%%%%%%%%%%%%%%%%
\subsection{Reheating contribution ($\mathcal{K}_{\rm reh}(k,N)$)}
\label{sec:Kreh}
%%%%%%%%%%%%%%%%%%%%%%%%%%%%%%%%%%%%%%%%%%%%%%%%%%%%%
We now evaluate the second term in Eq.~(\ref{eq:I2a}), corresponding to the contribution to the GW between $N_{\rm NR}$ and the end of reheating at $N_{\rm reh}$. Since the equation of state remains that of matter, we can reproduce the expression derived for $\mathcal{K}_{\rm NR}$, adjusting the
integration range and setting both $\mathcal{T}_X\to 1$ and $\mathcal{T}_{\rm M}\to 1$ (the Green's function is now evaluated only up to $N_{\rm reh}$):
\begin{align}\notag
\mathcal{K}_{\rm reh}(k,N) \;&=\;  \frac{1 }{4\sqrt{6}}  \left(\frac{k}{k_{\rm end}}\right)^3\left(\frac{a_{\rm end}H_{\rm end}}{a_{\rm reh}H_{\rm reh}}\right)   \int_{N_{\rm NR}}^{N_{\rm reh}} dN'\,\frac{\mathcal{G}_{\bk}(N_{\rm reh},N') }{(a(N')/a_{\rm end})^3} \left(\frac{a_{\rm end}H_{\rm end}}{a(N')H(N')}\right)^{2}\\ 
&\simeq\;  \frac{1}{4\sqrt{6}}   \left(\frac{k}{k_{\rm end}}\right)^3 \left(\frac{a_{\rm end}H_{\rm end}}{a_{\rm reh}H_{\rm reh}}\right)  \int_{N_{\rm NR}}^{N_{\rm reh}} dN'\,\mathcal{G}_{\bk}(N_{\rm reh},N') e^{-2(N'-N_{\rm end})}\,.
\end{align}
Integration of this expression for $T_{\rm reh}=10^{13}\,{\rm GeV}$ and $10^9\,{\rm GeV}$ yields the green curves shown in Fig.~\ref{fig:ks}. For modes that have not re-entered the horizon by the end of reheating, we observe the same scaling law as in the case of $\mathcal{K}_{\rm NR}$. The absence of
the transfer function $\mathcal{T}_X$ in the integrand allows a fully analytical estimate: applying the superhorizon Green's function~\eqref{eq:Gkrehapp}, we obtain
\begin{align} \notag
\mathcal{K}_{\rm reh}(k,N) \;&\simeq\; \frac{1}{12\sqrt{6}}\left(\frac{k}{k_{\rm end}}\right)^3 \left(\frac{a_{\rm end}H_{\rm end}}{a_{\rm reh}H_{\rm reh}}\right)e^{-2(N_{\rm NR}-N_{\rm end})}\\ \label{eq:Krehf}
&\simeq\; \frac{1}{24\sqrt{3}}\left(\frac{k}{k_{\rm end}}\right)^3 \left(\frac{\rho_{\rm end}}{\rho_{\rm reh}}\right)^{1/6}e^{-2(N_{\rm NR}-N_{\rm end})}\,. \qquad (k\ll k_{\rm reh})
\end{align}
As expected, $\mathcal{K}_{\rm reh}\propto k^3$ in this regime. The dependence on the energy density at the end of reheating implies $\mathcal{K}_{\rm reh}\propto T_{\rm reh}^{-2/3}$ (since $\rho_{\rm reh}\propto T_{\rm reh}^4$), which is reflected in the relative amplitudes of the smooth, leftmost portions of the green curves in Fig.~\ref{fig:ks}. This does not, however, imply that lower reheating temperatures produce a stronger GW signal. Owing to the blue tilt, the maximum
value of $\mathcal{K}_{\rm reh}$ for which the superhorizon approximation~\eqref{eq:Krehf} applies corresponds to $k\sim k_{\rm reh}$.  Evaluating at this scale, $\mathcal{K}_{\rm reh}(k_{\rm reh},N)\propto (k_{\rm reh}/k_{\rm end})^3\,T_{\rm reh}^{-2/3} \propto T_{\rm reh}^{4/3}$, so the superhorizon contribution is maximized at the highest 
possible reheating temperatures.

For modes that are already inside the horizon at the end of reheating ($k\gg k_{\rm reh}$), the numerical evaluation shows that $\mathcal{K}_{\rm reh}$ transitions to a shallower power law, $\mathcal{K}_{\rm reh}(k,N)\propto k$. The maximum amplitude is again controlled by $k_{\rm reh}$, and we conclude that the GW signal associated with $\mathcal{K}_{\rm reh}$ is maximized for large $T_{\rm reh}$, as is clearly seen in Fig.~\ref{fig:ks}.

%%%%%%%%%%%%%%%%%%%%%%%%%%%%%%%%%%%%%%%%%%%%%%%%%%%%%
\subsection{Radiation contribution ($\mathcal{K}_{\rm rad}(k,N)$)}
%%%%%%%%%%%%%%%%%%%%%%%%%%%%%%%%%%%%%%%%%%%%%%%%%%%%%
Continuing the evaluation of $\mathcal{I}(k,N)$, we turn to the late-time contribution $\mathcal{K}_{\rm rad}$, which is universal and depends only on the background dynamics (provided
$N_{\rm NR}<N_{\rm reh}$). From Eq.~\eqref{eq:I2a},
\begin{align}\notag
\mathcal{K}_{\rm rad}(k,N) \;&=\;  \frac{1}{4\sqrt{3}}  \left(\frac{k}{k_{\rm end}}\right)^3 \left(\frac{a_{\rm end}H_{\rm end}}{a(N)H(N)}\right)  \int_{N_{\rm reh}}^N dN'\,\frac{\mathcal{G}_{\bk}(N,N') }{(a(N')/a_{\rm end})^3} \left(\frac{a_{\rm end}H_{\rm end}}{a(N')H(N')}\right)^{2} \\ \notag
&=\;  \frac{1}{4\sqrt{3}} \left(\frac{k}{k_{\rm end}}\right)^3\left(\frac{a_{\rm end}H_{\rm end}}{a(N)H(N)}\right) \left(\frac{a_{\rm end}H_{\rm end}}{a_{\rm reh}H_{\rm reh}}\right)^2 \left(\frac{a_{\rm end}}{a_{\rm reh}}\right)^3   \int_{N_{\rm reh}}^N dN'\,\mathcal{G}_{\bk}(N,N') e^{-(N'-N_{\rm reh})} \\ \label{eq:Krad1}
&\equiv\;  \frac{1}{4\sqrt{3}} \left(\frac{k}{k_{\rm end}}\right)^3 \left(\frac{a_{\rm reh}H_{\rm reh}}{a(N)H(N)}\right) \left(\frac{a_{\rm end}H_{\rm end}}{a_{\rm reh}H_{\rm reh}}\right)^3 \left(\frac{a_{\rm end}}{a_{\rm reh}}\right)^3 \tilde{\gamma}(N;k) \,,
\end{align}
where in the second line we have substituted the redshift of the Hubble parameter in a radiation epoch ($H\propto a^{-2}$). The function $\widetilde\gamma(N;k)$ in the third line is defined by
the Green's function integral of the second line. The tilde distinguishes it from the analogous integral function in the case of isocurvature-sourced GWs~\cite{Garcia:2025wmu}. During radiation domination, the solution of Eq.~\eqref{eq:heqrad} yields the Green's function
\begin{align} \notag 
\mathcal{G}_{\bk}(N,N') &\;=\; \left(\frac{k_{\rm reh}}{k}\right)e^{-(N-N_{\rm reh})} \sin\left[\left(e^{N-N_{\rm reh}}-e^{N'-N_{\rm reh}}\right)\frac{k}{k_{\rm reh}}\right]\label{eq:greenrad}\\
&\;=\; \left(\frac{a(N)H(N)}{k}\right)\sin\left[\frac{k}{a(N)H(N)}-\frac{k}{a(N')H(N')}\right]\,.
\end{align}
This allows for a closed form expression for $\tilde{\gamma}(N;k)$ in terms of the trigonometric integral functions
\begin{align}\notag
\tilde{\gamma}(N;k) &\;=\; \left(\frac{k_{\rm reh}}{k}\right) e^{-(N-N_{\rm reh})} \Bigg\{\bigg[ \cos\left(\tfrac{k}{k_{\rm reh}}\right) + \left(\tfrac{k}{k_{\rm reh}}\right){\rm Si}\left(\tfrac{k}{k_{\rm reh}}\right) -  \left(\tfrac{k}{k_{\rm reh}}\right) {\rm Si}\left(e^{N-N_{\rm reh}}\left(\tfrac{k}{k_{\rm reh}}\right)\right)  \bigg]\\ \notag
&\qquad\qquad \times \sin\left(e^{N-N_{\rm reh}}\left(\tfrac{k}{k_{\rm reh}}\right)\right) - \bigg[ \sin\left(\tfrac{k}{k_{\rm reh}}\right) - \left(\tfrac{k}{k_{\rm reh}}\right){\rm Ci}\left(\tfrac{k}{k_{\rm reh}}\right)\\ 
&\qquad\qquad+ \left(\tfrac{k}{k_{\rm reh}}\right){\rm Ci}\left(e^{N-N_{\rm reh}}\left(\tfrac{k}{k_{\rm reh}}\right)\right)  \bigg]\cos\left(e^{N-N_{\rm reh}}\left(\tfrac{k}{k_{\rm reh}}\right)\right) \Bigg\}\\ \label{eq:gammaapp}
&\;\simeq\; \begin{cases}
1\,, & k/aH\ll 1\,,\\[5pt]
-\left(\dfrac{\sin(k/k_{\rm reh})}{k/k_{\rm reh}}\right)\left(\dfrac{\sin(k/a(N)H(N))}{k/a(N)H(N)}\right)\,, & k/aH\gg1 \,.
\end{cases}
\end{align}

As discussed above and shown explicitly in the next section, the GW signal is maximal for large reheating temperatures, corresponding to the curvaton-like scenario with an unstable spectator.  We can
express the background functions at the time of decay by noting that the spectator-to-radiation ratio is approximately preserved from $N_{\rm NR}$ through reheating: $(\rho_\chi/\rho_\phi)_{\rm NR}\simeq
(\rho_\chi/\rho_\phi)_{\rm reh} =(\rho_\chi/\rho_R)_{\rm reh}$. With $H\propto\rho_R^{1/2}\propto a^{-2}$ during radiation domination, this implies
\beq
\left(\frac{a_{\rm reh}H_{\rm reh}}{a(N)H(N)}\right)  \left(\frac{\rho_{\chi,{\rm NR}}}{H^2_{\rm NR} M_P^2}\right) \;=\; \frac{3}{2}f_{\chi}\,.
\eeq
Substitution into (\ref{eq:Krad1}) finally gives 
\begin{align} \notag
\mathcal{K}_{\rm rad}(k,N) \;&=\;  \frac{\sqrt{3}f_{\chi}}{8}\left(\frac{k}{k_{\rm end}}\right)^3 \left( \frac{ H_{\rm NR}^2M_P^2}{\rho_{\chi,{\rm NR}}} \right) \left(\frac{a_{\rm end}H_{\rm end}}{a_{\rm reh}H_{\rm reh}}\right)^3 \left(\frac{a_{\rm end}}{a_{\rm reh}}\right)^3  \tilde{\gamma}(N;k) \\
&=\; \frac{\sqrt{3} f_{\chi}}{8}\left(\frac{k}{k_{\rm end}}\right)^3 \left(\frac{\rho_{\rm reh}}{\rho_{\rm end}}\right)^{1/2} \left( \frac{ H_{\rm NR}^2M_P^2}{\rho_{\chi,{\rm NR}}} \right)  \tilde{\gamma}(N;k)\,.
\end{align}

Fig.~\ref{fig:ks} shows the evaluation of this expression for $f_\chi=1$ as the blue curves.  For modes that remain outside the horizon at the moment of decay, $k\ll a(N_{\rm d})H(N_{\rm d})$, the radiation contribution is smooth with $\mathcal{K}_{\rm rad}\propto k^3$. For $T_{\rm reh}=10^{13}$~GeV, the late-time dynamics in fact dominate the GW signal in this regime, with the contribution scaling as $\sim T_{\rm reh}^2$.  Nevertheless, owing to the blue tilt, the maximal amplitude is reached only for modes that have re-entered the horizon by the time of decay.  For those modes, the sub-horizon limit of $\widetilde\gamma$~\eqref{eq:gammaapp} produces an oscillatory behavior on top of a broken power law: $\mathcal{K}_{\rm rad}\propto k^2$ for $a(N_{\rm d})H(N_{\rm d})\ll k\ll k_{\rm reh}$, and $\mathcal{K}_{\rm rad}\propto k$ for $k\gg k_{\rm reh}$. Consequently, the largest amplitude of the radiation contribution to $\Omega_{{\rm GW},X}$ is obtained for the highest reheating temperatures. For both values of $T_{\rm reh}$ shown in Fig.~\ref{fig:ks}, $\mathcal{K}_{\rm rad}$ is the subdominant component.

%%%%%%%%%%%%%%%%%%%%%%%%%%%%%%%%%%%%%%%%%%
\section{Gravitational Wave Signals}
\label{sec:gwsignals}
%%%%%%%%%%%%%%%%%%%%%%%%%%%%%%%%%%%%%%%%%%

%%%%%%%%%%%%%%%%%%%%%%%%%%%%%%%%%%%%%%%%%%
\subsection{Detection Prospects}
\label{sec:gwmissions}
%%%%%%%%%%%%%%%%%%%%%%%%%%%%%%%%%%%%%%%%%%

The stochastic GW background generated by spectator fields peaks at ultra high frequencies $f\sim 10^3 - 10^9$~Hz, set by the comoving Hubble scale at the end of inflation.  This places the signal well above the sensitivity bands of all current and planned interferometric GW detectors, and below the integrated energy density bounds from BBN and CMB constraints on $\Delta N_{\rm eff}$~\cite{Maggiore:1999vm,Luo:2020fdt, Cyburt:2015mya,Planck:2018vyg}. The predicted signal is therefore currently unconstrained by observations.

Ground-based interferometers, such as Advanced LIGO/Virgo~\cite{LIGOScientific:2019lzm}, Einstein
Telescope~\cite{Punturo:2010zz}, and Cosmic Explorer~\cite{Reitze:2019iox}, operate at
$f\sim 1 - 10^4$~Hz. Space-based missions (LISA~\cite{LISA:2017pwj}, DECIGO and
Ultimate-DECIGO~\cite{Yagi:2011wg}, BBO~\cite{Crowder:2005nr}) and atom interferometers
(AION~\cite{Badurina:2021rgt}, AEDGE~\cite{AEDGE:2019nxb}) cover the
$10^{-4}-10$~Hz range, while proposed observatories such as $\mu$-Ares~\cite{Sesana:2019vho} would explore the $\mu$Hz window. In the nHz regime, pulsar timing arrays including
NANOGrav~\cite{NANOGrav:2023hvm} and EPTA~\cite{EPTA:2015qep} provide sensitivity at
    $10^{-9}-10^{-7}$~Hz. None of these experiments, however, reach the ultra high-frequency regime where the spectator-induced signal peaks. The steep infrared tail ($\Omega_{\rm GW}h^2\propto f^5$) falls well below their sensitivity floors for the benchmark parameters considered in this work. 

Direct detection of these signals would require ultra high-frequency GW detectors, an area of active experimental development~\cite{Aggarwal:2020olq,Herman:2022fau,Ringwald:2022xif}. While current concepts based on resonant cavities and other novel techniques have not yet reached the sensitivities needed to probe the parameter space of interest, the strong theoretical motivation
provided by spectator-induced signals, with predicted amplitudes as large as $\Omega_{\rm GW}h^2\sim 10^{-11}$, underscores the importance of continued progress in this frontier.

%%%%%%%%%%%%%%%%%%%%%%%%%%%%%%%%%%%%%%%%%%
\subsection{Curvaton-like scenarios with decaying spectator fields}
\label{sec:curvatondecay}
%%%%%%%%%%%%%%%%%%%%%%%%%%%%%%%%%%%%%%%%%%

The factorization of the GW spectrum into the momentum-dependent integral $g(k)$ and the time-dependent factor $\mathcal{I}(k,N)$, as expressed in Eq.~\eqref{eq:OGWfact}, allows for a
straightforward (numerical) computation of the signal.

As discussed in detail in the previous section, our main regime of interest corresponds to curvaton-like scenarios with a late decay ($f_\chi \simeq 1$), large inflaton-spectator couplings
($\sigma/\lambda\gg 1$), and high reheating temperatures. Under these assumptions, the dominant contribution to $\mathcal{I}(k,N)$ comes from the early-time component $\mathcal{K}_{\rm NR}(k,N)$.
Applying the approximations~\eqref{eq:gk_analytic} and~\eqref{eq:Krehf}, we can analytically estimate the amplitude of the GW signal for scales that remain outside the horizon by the end
of reheating ($k\ll k_{\rm reh}$):
\begin{align} \notag
\Omega_{{\rm GW},X} \; &\simeq \; g(k) \mathcal{K}_{\rm NR}^{2}(k,N)  \\  
\label{eq:GWnr}
&=\; \frac{4}{2835}\left(\frac{k}{k_{\rm end}}\right)^5 \left(\frac{\mathcal{K}_{\rm NR}}{\mathcal{K}_{\rm reh}}\right)^2 \left(\frac{k_{\rm peak}}{k_{\rm end}}\right)^7 \left(\frac{\rho_{\rm end}}{\rho_{\rm reh}}\right)^{1/3} \left(\frac{\widetilde\Delta_X^2(k)}{k^3}\right)^2 e^{-4(N_{\rm NR}-N_{\rm end})} \,.
\end{align}
Here we have used the proportionality $\mathcal{K}_{\rm NR}\propto\mathcal{K}_{\rm reh}$, where the
numerical prefactor must be determined numerically due to the nontrivial evolution of the spectator spectrum during the resonant regime (see Fig.~\ref{fig:PX_T}).  For smaller scales, the shape of
the spectrum depends on the coupling strengths, the mass of $\chi$, and the expansion history parameterized by $T_{\rm reh}$.

%%%%%%%%%%%%%%%%%%%%%%%%%%%%%%%%%%%%%%%%%%
\subsubsection{No self-interaction ($\lambda_{\chi} = 0$)}
\label{sec:GW_lchi0}
%%%%%%%%%%%%%%%%%%%%%%%%%%%%%%%%%%%%%%%%%%

\begin{figure*}[!t]
\centering
    \includegraphics[width=0.7\textwidth]{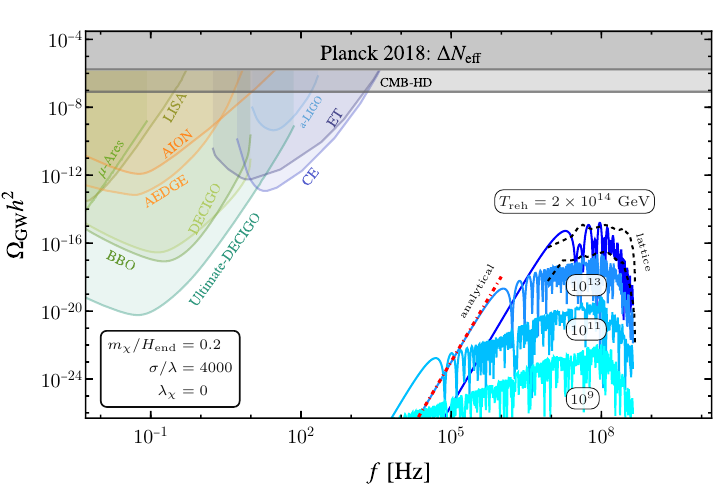}
    \caption{Gravitational wave spectrum sourced by the fluctuations of an unstable scalar spectator field $\chi$ ($f_\chi(t_{\rm d})=1$) with $m_\chi/H_{\rm end}=0.2$, $\sigma/\lambda=4000$, and vanishing self-interaction $\lambda_\chi=0$. The mass and the coupling of $\chi$ are fixed to values that allow for strong parametric resonance during reheating, and a rapid transition to the adiabatic, non-relativistic regime for $\rho_{\chi}$. The reheating temperature is varied from $T_{\rm reh}=10^9$ to $2\times 10^{14}$~GeV (solid colored curves: numerical evaluation of $\Omega_{{\rm GW},X}$ in the Hartree approximation using Eq.~\eqref{eq:OGWfact}). The black dashed curves show the spectrum computed with \textsc{CosmoLattice}~\cite{Figueroa:2020rrl,Figueroa:2021yhd}
    for the two largest reheating temperatures. The red dot-dashed line shows the analytical approximation~\eqref{eq:GWnr}.  Shaded regions show the sensitivity curves of current and proposed GW observatories, the gray band denotes the \textit{Planck}~2018 $\Delta N_{\rm eff}$ bound, and the darker line within it marks the projected CMB-HD sensitivity.}
    \label{fig:GW_Treh}
\end{figure*}

Figure~\ref{fig:GW_Treh} displays the GW spectrum for a spectator with $m_\chi/H_{\rm end}=0.2$, a large portal coupling $\sigma/\lambda=4000$, and vanishing self-interaction $\lambda_\chi=0$. The reheating temperature is varied discretely from $T_{\rm reh}=2\times10^{14}$~GeV down to $10^7$~GeV.  The largest value corresponds to {\em almost} instantaneous reheating: the assumption of a transition to radiation domination immediately after the maximum of the broad parametric resonance has been reached, $a_{\rm reh}\simeq 20$ (see Fig.~\ref{fig:rhos}). The rest of the reheating temperatures have been chosen so that the non-relativistic regime for $\chi$ is reached before the end of reheating ($N_{\rm NR}<N_{\rm reh}$). To convert the momentum dependence of the spectrum into the frequency domain we have used the relation 
\begin{align}
    f \; = \; \frac{1}{2\pi}\left(\frac{k_{\rm end}}{\rm Hz}\right)\left(\frac{k}{k_{\rm end}}\right)\,{\rm Hz} \; \simeq\; 2.43\times 10^7\, \left(\frac{k}{k_{\rm end}}\right)\,{\rm Hz}\;\simeq\; 2.43\times 10^7\,\left(\frac{k}{1.6\times 10^{22}\ {\rm Mpc}^{-1}}\right)\, {\rm Hz}\,.
\end{align}

For superhorizon modes during reheating ($k\ll k_{\rm reh}$), the approximation~\eqref{eq:GWnr} predicts larger signals as the reheating temperature is decreased, $\Omega_{{\rm GW},X}\propto T_{\rm reh}^{-4/3}$. This behavior can be clearly appreciated in Fig.~\ref{fig:GW_Treh}, where the frequency corresponding to the horizon size at the end of reheating coincides with the first peak of the spectrum to the left. With $\mathcal{K}_{\rm NR}\simeq 106\,\mathcal{K}_{\rm reh}$ for the present choice of couplings, and $k_{\rm peak}$, $\widetilde\Delta_X^2$ readily available from Fig.~\ref{fig:PX_T}, the analytical estimate~\eqref{eq:GWnr} can be immediately compared to the exact solution in this range of frequencies.  The result, shown as the red dot-dashed curve for $T_{\rm reh}=10^{13}$~GeV, confirms the accuracy of this approximation. The same general trend is followed by the spectrum with $T_{\rm reh}=2\times 10^{14}$~GeV, although we do not show an analytical approximation for it, as the field spectrum is still evolving non-trivially when reheating is completed, requiring the full numerical evaluation of $\mathcal{K}_{\rm NR}$ (see Fig.~\ref{fig:PX_T}). Following our discussion in Section~\ref{sec:Kreh}, although at very low frequencies a smaller reheating temperature enhances the amplitude of the GW spectrum, a larger $T_{\rm reh}$ extends this featureless portion of the spectrum to higher frequencies, and consequently larger amplitudes, such that $\Omega_{{\rm GW},X}(k_{\rm reh})\propto T_{\rm reh}^{8/3}$. Therefore, the amplitude of the GW signal is maximal for a rapidly decaying inflaton. It is also worth noting that, for the large value of $\sigma$ considered in this case, the suppression of growth for superhorizon $\chi$ modes during inflation, and as a consequence the steep blue tilt $\Omega_{{\rm GW},X}\propto k^5 \sim f^5$, pushes the signal to ultra high frequencies well above the sensitivity bands of all current and planned interferometric GW observatories, as shown by the lack of intersection between the spectra and the collection of sensitivity curves in Fig.~\ref{fig:GW_Treh}. Nevertheless, the predicted amplitudes, reaching $\Omega_{\rm GW}h^2\sim 10^{-11}$ near the spectral peak for $T_{\rm reh}=2\times 10^{14}$~GeV, provide strong motivation for the
development of ultra high-frequency GW detection techniques~\cite{Aggarwal:2020olq,Herman:2022fau,Ringwald:2022xif}.

At smaller scales, with $k\gg k_{\rm reh}$, the GW spectrum becomes more challenging to approximate analytically. For $T_{\rm reh}\gtrsim 10^{13}$~GeV, the dominant contribution from background dynamics to the GW spectrum continues to be $\mathcal{K}_{\rm NR}$, as shown in Fig.~\ref{fig:ks}.  Therefore, the signal can be approximated from Eq.~\eqref{eq:GWnr} by adjusting the tilt, with $\Omega_{{\rm GW},X}\propto k$ in this regime, up to $k\lesssim k_{\rm end}$, which corresponds to $f\lesssim f_{\rm end}\simeq 2\times 10^7$~Hz. The oscillatory features are a consequence of the oscillation of the Green's function for the tensor modes as they re-enter the horizon during the matter-dominated reheating epoch. For even larger frequencies the analytical control is lost, as the prefactor $g(k)$ ceases to behave like a power law and requires numerical evaluation, as shown in Fig.~\ref{fig:gk}.  Furthermore, the factorization of the field spectrum~\eqref{eq:TX} into time-dependent and $k$-dependent functions ceases to be valid (see Fig.~\ref{fig:PX_T}). Therefore, for $f\gtrsim f_{\rm end}$ we do not expect our computation based on Eq.~\eqref{eq:OGWfact} to accurately reflect the amplitude of the GW spectrum.  However, this regime of small-scale inhomogeneities can be explored by means of lattice methods.  The black dashed curves in Fig.~\ref{fig:GW_Treh} shows the result of evaluating the GW spectrum with \textsc{CosmoLattice} for subhorizon, and mildly superhorizon, modes during reheating.  Despite the breakdown of the factorized approximation, there is good agreement between the lattice and Hartree results. The two approaches are in this sense complementary: the lattice captures the full nonlinear dynamics near and above the spectral peak, while the Hartree calculation extends the prediction to deeply superhorizon scales that would be prohibitively expensive to simulate on the lattice.

For $T_{\rm reh}< 10^{13}$~GeV, the previous analysis is mostly unchanged, despite $\mathcal{K}_{\rm NR}<\mathcal{K}_{\rm reh}$. As we discussed previously, both contributions have a similar $k$-dependence.  We can safely conclude that the amplitude is an overall decreasing function of $T_{\rm reh}$ for $f\sim f_{\rm end}$. To finish the discussion concerning Fig.~\ref{fig:GW_Treh}, we note that despite our initial assumption of $f_\chi=1$, for all cases shown the late-time contribution $\mathcal{K}_{\rm rad}$ is negligible, and therefore the results are applicable also for any $f_\chi\leq 1$, provided the decay occurs sufficiently late after reheating not to compromise the validity of Eq.~\eqref{eq:TRav}.\par\bigskip

\begin{figure*}[!t]
\centering
    \includegraphics[width=0.7 \textwidth]{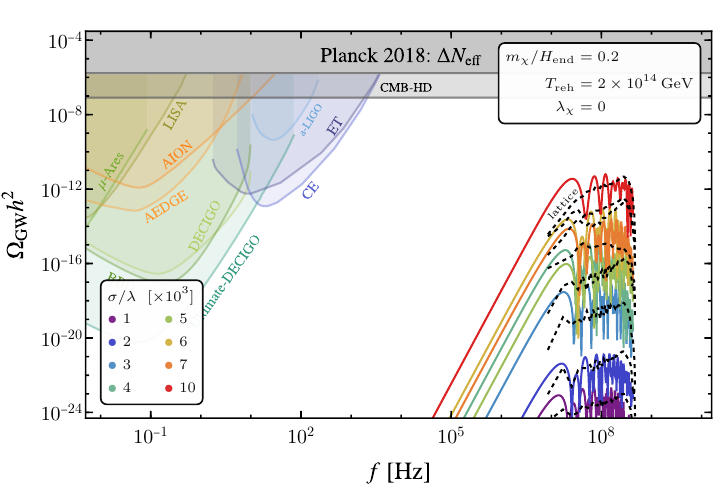}
    \includegraphics[width=0.7 \textwidth]{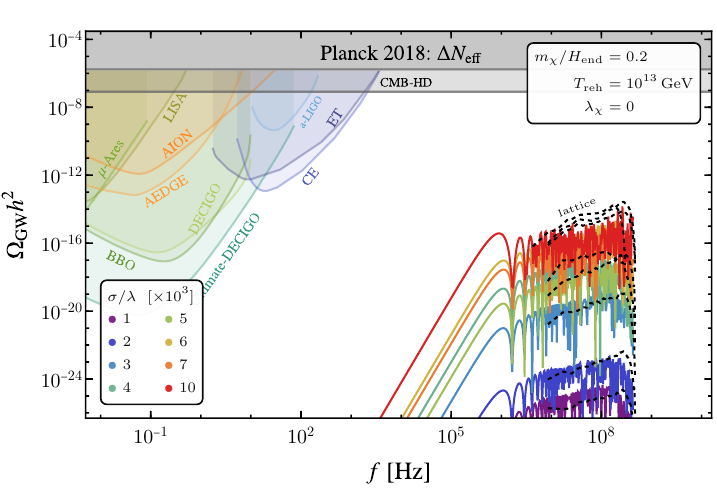}
    \caption{Gravitational wave spectrum sourced by the fluctuations of an unstable scalar spectator field $\chi$ ($f_\chi(t_{\rm d})=1$), with $m_\chi/H_{\rm end}=0.2$, $\lambda_\chi=0$, and $T_{\rm reh}=2\times 10^{14}\,{\rm GeV}$ (top) and $T_{\rm reh}=10^{13}\,{\rm GeV}$ (bottom). The inflaton-spectator coupling is varied over $\sigma/\lambda=10^3$--$10^4$ (colored solid curves). For each coupling, the corresponding GW spectrum computed with \textsc{CosmoLattice} is shown as a black dashed curve.}
    \label{fig:GW_sigma}
\end{figure*}

Fig.~\ref{fig:GW_sigma} presents the results of a discrete variation of the coupling strength between the $\phi$ and $\chi$ fields in the strong-coupling regime, fixing the spectator mass and
other couplings to the values used in Fig.~\ref{fig:GW_Treh}, with $T_{\rm reh}=2\times 10^{14}\,{\rm GeV}$ (top) and $T_{\rm reh}=10^{13}\,{\rm GeV}$ (bottom).  For all but the largest coupling considered, the analysis performed previously for $\sigma/\lambda=4000$ applies. In
particular, the approximation~\eqref{eq:GWnr} provides an accurate estimate of $\Omega_{{\rm GW},X}$ for $k\ll k_{\rm reh}$, while at smaller scales the spectral envelope follows $\Omega_{{\rm GW},X}\propto k \sim f$ up to $k_{\rm end}$, beyond which one must rely on the full numerical computation in the Hartree approximation. The main feature to be observed in the figure is the
dependence of the GW amplitude on the effective coupling $\sigma/\lambda$. The hierarchy of amplitudes follows closely the corresponding spectator energy densities, as shown in the left panel
of Fig.~\ref{fig:rhos_sl}: the monotonic dependence on $\sigma$ is lost for large couplings, owing to the complex structure of the broad parametric resonance. 

For comparison, Fig.~\ref{fig:GW_sigma} also shows the GW spectra computed using lattice methods (black dashed curves). We note good agreement with the Hartree curves, in the envelope sense, for the
four smallest couplings. As $\sigma/\lambda$ is increased, the presence of inflaton fluctuations and rescatterings can no longer be ignored, and we observe a degradation of the agreement between the
Hartree and lattice results for the largest couplings. This is consistent with the comparison of the spectator energy densities in Fig.~\ref{fig:rhos_sl}, where the Hartree approximation increasingly
underestimates the spectator abundance at large
$\sigma/\lambda$. For the sake of completeness, we have also included the result of applying our analysis for a coupling for which a strong backreaction in the inflaton dynamics is present ($\sigma/\lambda=10^4$). In the Hartree approximation, we account for this effect in the homogeneous limit; (\ref{eq:kg_conf}) is replaced by
\beq
\phi'' + 2\mathcal{H}\phi' + a^2 (V_{,\phi}+ \sigma\langle \chi^2\rangle \phi) \; = \; 0 \,.
\eeq
A more complete description would include the effect of mode-mode couplings and the fragmentation of the inflaton, potentially invalidating our results. Nevertheless, we observe that the more accurate lattice computation, which incorporates non-linear effects, is still in relatively good agreement with the mean-field approximation.\par\bigskip

\begin{figure*}[!t]
\centering
    \includegraphics[width=0.7\textwidth]{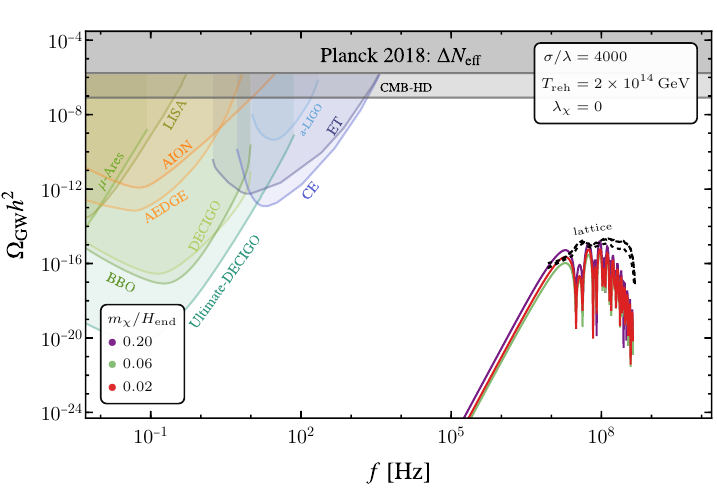}
    \includegraphics[width=0.7\textwidth]{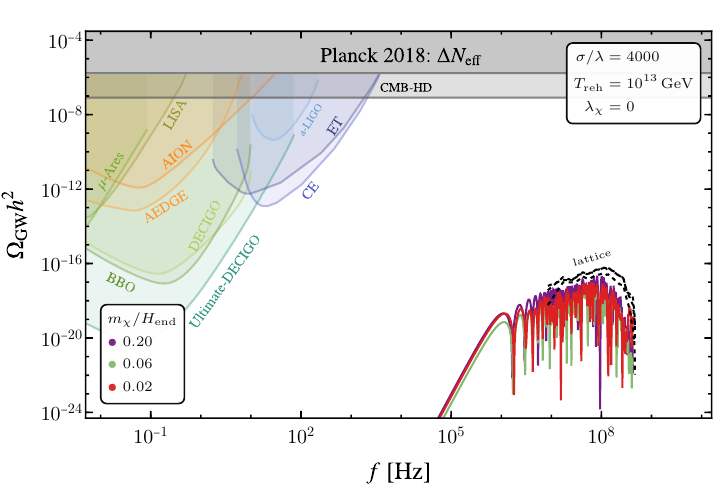}
    \caption{Mass dependence of the gravitational wave spectrum sourced by a fluctuating unstable scalar spectator field $\chi$, with $\sigma/\lambda=4000$, $\lambda_\chi=0$, and $T_{\rm reh}=2\times 10^{14}\,{\rm GeV}$ (top) and $T_{\rm reh}=10^{13}\,{\rm GeV}$ (bottom) held fixed. Three values of $m_\chi/H_{\rm end}$, spanning an order of magnitude, are shown. For each mass, the GW spectrum computed with \textsc{CosmoLattice} is shown as a black dashed curve.}
    \label{fig:GW_mass}
\end{figure*}

Figure~\ref{fig:GW_mass} shows the dependence of $\Omega_{{\rm GW},X}$ on the mass of the spectator field, for three values within an order of magnitude of each other. This range was selected so that the two larger masses satisfy the condition $N_{\rm NR}<N_{\rm reh}$ for $T_{\rm reh}\lesssim 10^{13}\,{\rm GeV}$ assumed in our previous assessment of the GW spectra. For the smallest mass this condition is weakly violated, and we therefore rely on a fully numerical computation, made more expensive by the need to extend the integration past the end of reheating. The main result that can be deduced from the figure is that the GW spectrum is insensitive to the mass of $\chi$. This weak sensitivity is due to the large coupling $\sigma$, which determines the strength of the parametric resonance and thus the growth of the field power spectrum. For $\sigma/\lambda\gtrsim 10^3$, the effective mass~\eqref{eq:meff_total} is dominated by $m_{\chi,\rm eff}^2\simeq\sigma\phi^2$ during the early stages of reheating (as long as $m_\chi\ll H_{\rm end}$). The only remaining dependence of $\Omega_{{\rm GW},X}$ on $m_\chi$ therefore enters through $N_{\rm NR}$, whose implicit appearance in the prefactor $(\mathcal{K}_{\rm NR}/\mathcal{K}_{\rm reh})$ in Eq.~\eqref{eq:GWnr} makes a precise analytical estimate difficult.

In Fig.~\ref{fig:GW_mass} we also show the corresponding lattice computation for the GW spectra. The smallest two masses lead to almost indistinguishable curves, while the smallest mass gives a spectrum with a slightly larger amplitude, reproducing the aforementioned behavior.   
%%%%%%%%%%%%%%%%%%%%%%%%%%%%%%%%%%%%%%%%%%
\subsubsection{Nonvanishing self-interaction ($\lambda_\chi\neq 0$)}
\label{sec:GW_lchi}
%%%%%%%%%%%%%%%%%%%%%%%%%%%%%%%%%%%%%%%%%%

We now briefly discuss the GW spectrum sourced by a spectator field with a sizable quartic self-interaction. As specified in Eq.~\eqref{eq:meff_total}, we treat the self-interaction within the
Hartree approximation, where it contributes to the effective mass of $\chi$ through the expectation value $\langle\chi^2\rangle$, rather than as a convolution coupling modes with different wavenumbers. We
expect this approximation to be valid for sufficiently small values of $\lambda_\chi$, and determine its range of validity by comparison with lattice computations, which are performed in configuration
space and are therefore free of such limitations.

\begin{figure*}[!t]
\centering
    \includegraphics[width=0.7\textwidth]{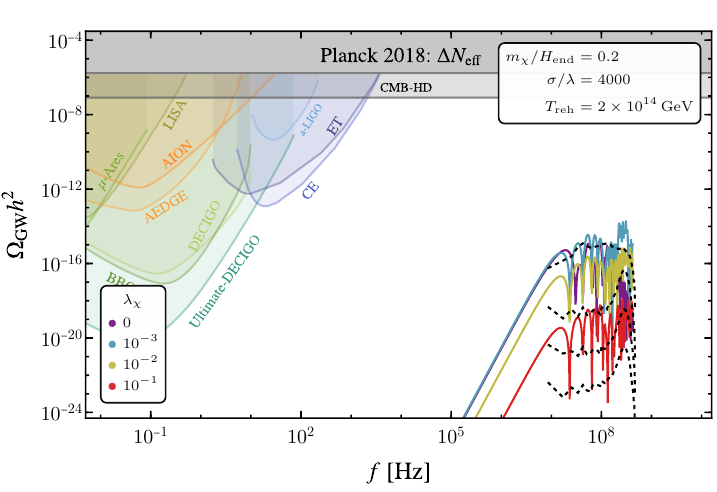}
    \includegraphics[width=0.7\textwidth]{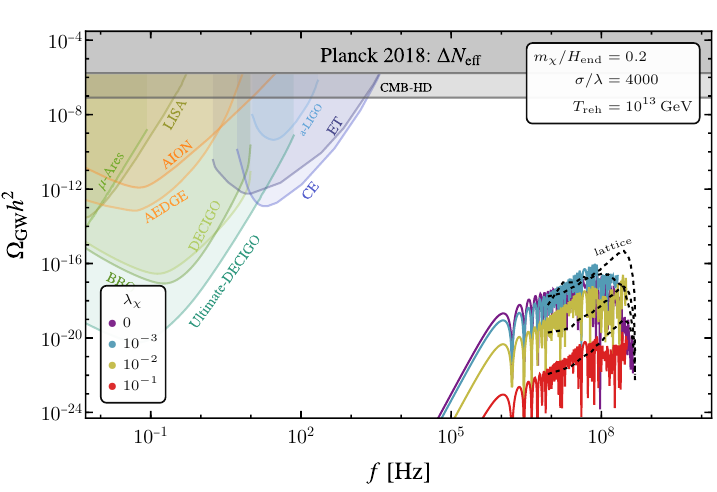}
    \caption{Gravitational wave spectrum sourced by a  self-interacting scalar spectator field $\chi$, with $m_\chi/H_{\rm end}=0.2$, $\sigma/\lambda=4000$, and $T_{\rm reh}=2\times 10^{14}$~GeV (top) and $T_{\rm reh}=10^{13}$~GeV (bottom). The self-interaction coupling $\lambda_\chi$ is varied as indicated (colored solid curves:
    Hartree approximation). Black dashed curves show the corresponding spectra computed with
    \textsc{CosmoLattice}.}
    \label{fig:GW_self}
\end{figure*}

Fig.~\ref{fig:GW_self} shows the result of varying the magnitude of the self-interaction coupling $\lambda_\chi$, for the same values considered in Fig.~\ref{fig:rhos_sl}. The dependence of the GW amplitude on $\lambda_\chi$ is non-monotonic.  For a small self-coupling ($\lambda_\chi=10^{-3}$), the GW spectrum is \emph{enhanced} relative to the $\lambda_\chi=0$ case for UV modes. This enhacement is present in the Hartree approximation for the two reheating temperatures shown. For larger values ($\lambda_\chi=10^{-2},\,10^{-1}$) the amplitude is progressively suppressed. This behavior, which is mirrored in the spectator energy densities (right panel of Fig.~\ref{fig:rhos_sl}), can be understood as a competition between two effects of the quartic vertex. At small $\lambda_\chi$, mode-mode rescattering induced by the $\lambda_\chi\chi^4$ interaction provides additional channels for redistributing spectral weight, effectively broadening the resonance band and enhancing the overall particle production before the Hartree mass $\frac{\lambda_\chi}{2}\langle\chi^2\rangle$ becomes large enough to detune the parametric resonance~\cite{Figueroa:2023oxc, Micha:2002ey, Micha:2004bv}. At larger $\lambda_\chi$, the Hartree backreaction kicks in earlier, shutting off the resonance before the spectrum can grow to its full extent, and the suppression wins.

The lattice spectra (black dashed curves) show a more complicated behavior of the GW spectra. For a sudden inflaton decay, with $T_{\rm reh}=2\times 10^{14}\,{\rm GeV}$, an inverse dependence of the amplitude of the spectra on the coupling $\lambda_{\chi}$ is found at all frequencies. The agreement with the Hartree approximation is poor in this case. Remarkably, however, for a longer duration of reheating, with $T_{\rm reh}=10^{13}\,{\rm GeV}$, the agreement is improved. In this case we confirm both the non-monotonic trend and the overall amplitude hierarchy.  In addition to the amplitude variation, the lattice results reveal enhanced UV power at high frequencies for nonvanishing $\lambda_\chi$, arising from the turbulent cascade of energy to
short wavelengths through $2\to 2$ scattering processes that are absent in the Hartree
approximation~\cite{Figueroa:2023oxc,Micha:2002ey,Micha:2004bv}.

%%%%%%%%%%%%%%%%%%%%%%%%%%%%%%%%%%%%%%%%%%
\subsection{Gravitational Waves with Quartic Inflaton Potential}
\label{sec:quarticGWs}
%%%%%%%%%%%%%%%%%%%%%%%%%%%%%%%%%%%%%%%%%%

Throughout this work we have focused on the quadratic T-model
potential~\eqref{eq:tmodel_potential}, for which the inflaton
oscillates in a quadratic minimum and the post-inflationary
background is matter-dominated ($w_{\rm eff}\simeq 0$) until
reheating.  In this regime, the GW energy density redshifts faster
than radiation ($\rho_{\rm GW}\propto a^{-4}$ versus
$\rho_\phi\propto a^{-3}$), so $\Omega_{\rm GW}$ is diluted
relative to the background during the matter-dominated phase.  This
dilution is absent if the inflaton potential has a quartic minimum,
for which $w_{\rm eff}\simeq 1/3$ and both the GW energy density
and the background scale as $a^{-4}$.  The GW signal is therefore
expected to be significantly enhanced in this
case~\cite{Easther:2006gt,Easther:2006vd,Easther:2007vj,
Khlebnikov:1997di,Dufaux:2007pt,Figueroa:2017vfa}.

A natural realization within the $\alpha$-attractor framework is the
quartic T-model,
\begin{equation}
\label{eq:tmodel_potentialquartic}
    V(\phi)
    \;=\;
    \lambda M_P^4
    \left[\sqrt{6}\tanh\!\left(\frac{\phi}{\sqrt{6}\,M_P}\right)
    \right]^4,
\end{equation}
which shares the same plateau behavior at large field values as Eq.~\eqref{eq:tmodel_potential} but reduces to $V\simeq\lambda\phi^4$ near the origin, yielding a radiation-like equation of state during inflaton oscillations.

A full exploration of the quartic T-model parameter space, including the dependence on $\sigma/\lambda$, $\lambda_\chi$, and the spectator mass, as well as a semi-analytical framework analogous to the $\mathcal{K}$ decomposition developed for the quadratic case, is left for future work. We note that lattice studies of GW production from preheating with quartic inflaton potentials have a long history~\cite{Khlebnikov:1997di,Easther:2006gt,Easther:2006vd,
Easther:2007vj,Dufaux:2007pt,Figueroa:2017vfa}, and our results for the spectator-sourced signal are qualitatively consistent with the enhanced amplitudes found in those works for the inflaton-sourced case.

%%%%%%%%%%%%%%%%%%%%%%%%%%%%%%%%%%%%%%%%%%
\section{Conclusions}
\label{sec:conclusions}
%%%%%%%%%%%%%%%%%%%%%%%%%%%%%%%%%%%%%%%%%%
In this work we have developed a comprehensive framework for computing the stochastic gravitational wave background sourced by scalar-induced tensor perturbations from a spectator field $\chi$ coupled to the inflaton through a portal interaction $\sigma\phi^2\chi^2$. We have focused on the large portal coupling regime ($\sigma/\lambda\gg 1$), in which parametric amplification of spectator fluctuations during reheating, rather than purely gravitational particle production, is the dominant mechanism for
building up the spectator power spectrum that sources the GW signal. Our main results and conclusions are as follows.

We have shown that purely gravitational particle production is insufficient to generate a detectable GW signal in this class of models: the resulting spectator abundance is too small for any reheating temperature consistent with dark matter overproduction constraints. The large portal coupling $\sigma$ qualitatively changes the picture by inducing broad parametric resonance in the spectator mode equation during inflaton oscillations, amplifying the spectator power spectrum by up to $\sim 15$ orders of magnitude over $\sim 5$~$e$-folds of reheating (Fig.~\ref{fig:PX_T}). Simultaneously, the heavy effective mass $\sigma\phi_*^2\gg H_*^2$ during inflation suppresses superhorizon fluctuations, yielding a maximally blue-tilted spectrum ($\widetilde\Delta_X^2\propto k^3$) that comfortably satisfies CMB isocurvature constraints while concentrating the spectral weight at small scales where the GW
signal is largest. The quartic self-interaction $\lambda_\chi$ regulates the amplification through the Hartree backreaction mass $\frac{\lambda_\chi}{2}\langle\chi^2\rangle$, with the ratio $\sigma/\lambda$ governing the overall efficiency of the process.

We have derived a master formula for the GW power spectrum (Eq.~\eqref{eq:Deltah_master}) from first principles, including the vacuum subtraction $|X_q|^2-1/(2\omega_q)$ in each leg of the Wick
contraction, which we have verified numerically to be essential for ultraviolet convergence of the momentum integral. Exploiting the transfer function factorization of the spectator power spectrum in
the infrared, the master formula separates into a spectral integral $g(k)$ encoding the frozen spectator power spectrum and a time-dependent factor $\mathcal{I}(k,N)$ capturing the build-up of
the GW signal through the post-inflationary expansion history. The spectral integral admits an analytic $k^{-1}$ approximation in the infrared (Eq.~\eqref{eq:gk_analytic}), validated against the full
numerical evaluation (Fig.~\ref{fig:gk}).

We have decomposed the time-dependent factor into three physically distinct contributions: $\mathcal{K}_{\rm NR}$ (parametric amplification phase), $\mathcal{K}_{\rm reh}$ (adiabatic reheating),
and $\mathcal{K}_{\rm rad}$ (radiation domination), and derived analytical expressions for the tensor transfer functions in the matter- and radiation-dominated eras. For the curvaton-like
scenario with large $T_{\rm reh}$, the early-time contribution $\mathcal{K}_{\rm NR}$ dominates. This decomposition yields the analytical estimate~\eqref{eq:GWnr} for superhorizon modes, which
reproduces the full numerical result with excellent accuracy (Fig.~\ref{fig:GW_Treh}).

The GW amplitude depends sensitively on the reheating temperature. While the superhorizon spectrum scales as $\Omega_{{\rm GW},X}\propto T_{\rm reh}^{-4/3}$ at fixed $k$, the blue tilt extends the signal to higher frequencies for larger $T_{\rm reh}$, so that the peak amplitude scales as $\Omega_{{\rm GW},X}(k_{\rm reh})\propto T_{\rm reh}^{8/3}$. The GW signal is therefore maximized for the highest possible reheating temperatures. For $T_{\rm reh}=2 \times 10^{14}$~GeV and $\sigma/\lambda=10^4$, the peak amplitude reaches $\Omega_{\rm GW}h^2\sim 10^{-11}$ at frequencies $f\sim 10^7$--$10^8$~Hz.

We have validated the Hartree computation against full nonlinear lattice simulations performed with
\textsc{CosmoLattice}~\cite{Figueroa:2020rrl,Figueroa:2021yhd}. The agreement between the two approaches is excellent for the spectator power spectrum at $k\lesssim k_{\rm end}$ (Fig.~\ref{fig:PX_T}) and for the GW spectrum across the entire frequency range accessible to both methods (Figs.~\ref{fig:GW_Treh}-\ref{fig:GW_self}). The two approaches are complementary: the lattice captures the full nonlinear dynamics, including mode-mode rescattering, inflaton fragmentation, and UV cascades from the quartic self-interaction, while the Hartree calculation extends the prediction to deeply superhorizon scales
that would be prohibitively expensive to simulate on the lattice. Discrepancies emerge at large $\sigma/\lambda$ and large $\lambda_\chi$, where nonlinear effects beyond the mean-field approximation become significant, but the overall amplitude and spectral shape remain in qualitative agreement.

The GW spectrum depends only weakly on the bare spectator mass $m_\chi$ (a factor of $\sim 2$ across an order of magnitude in $m_\chi$), since the effective mass during the resonant phase is dominated by the portal contribution $\sigma\phi^2$. By contrast, the amplitude varies by many orders of magnitude across the range $\sigma/\lambda=10^3 - 10^4$ (Fig.~\ref{fig:GW_sigma}), closely tracking the spectator energy density produced during reheating. A nonvanishing quartic self-interaction $\lambda_\chi$ suppresses the signal by hastening the Hartree backreaction that detunes the resonance (Fig.~\ref{fig:GW_self}), while on the lattice it additionally generates UV spectral weight through mode-mode rescattering.

The steep infrared scaling $\Omega_{{\rm GW},X}\propto f^5$, inherited from the white-noise spectator spectrum through the convolution~\eqref{eq:gkint}, pushes the signal to ultra high frequencies far above the sensitivity bands of all current and planned GW interferometers.  The predicted amplitudes also lie below the integrated energy density bounds from BBN and CMB constraints on $\Delta N_{\rm eff}$.  The signal is therefore currently unconstrained by observations. Direct detection would require ultra high-frequency GW detectors based on resonant cavities or other novel techniques~\cite{Aggarwal:2020olq,Herman:2022fau,Ringwald:2022xif}, for which the predicted amplitudes of up to $\Omega_{\rm GW}h^2\sim 10^{-11}$ provide strong theoretical motivation.

Several directions for future work are suggested by our results. First, extending the analysis to include the metric-sourced ($\Phi\Phi$) contribution to the GW spectrum, which probes larger
scales and could potentially reach the sensitivity bands of space-based interferometers, would provide a more complete picture of the observational signatures. Second, a systematic exploration of
the full $(m_\chi,\,\sigma,\,\lambda_\chi,\,T_{\rm reh})$ parameter space, including the stable dark matter scenario, would map out the regions accessible to future $\Delta N_{\rm eff}$ measurements
and ultra high-frequency detectors. Third, the role of nonminimal gravitational couplings ($\xi\chi^2 R$), which can enhance the spectator power spectrum through tachyonic instabilities~\cite{Figueroa:2023oxc,Bettoni:2024ixe}, deserves further investigation as a mechanism for boosting the GW signal into observable ranges. Finally, a detailed comparison of the Hartree and
lattice approaches in the strong backreaction regime would sharpen the theoretical predictions for the largest portal couplings, where inflaton fragmentation and turbulent thermalization become important.

The framework developed in this work demonstrates that spectator scalar fields with portal couplings to the inflaton are a robust and well-motivated source of stochastic gravitational waves in the early
universe. The interplay of parametric amplification, Hartree backreaction, and the post-inflationary expansion history imprints distinctive spectral features that encode information about the
reheating dynamics and the spectator sector, physics that is otherwise inaccessible to conventional cosmological probes. While the predicted signals lie beyond the reach of current experiments,
they define a clear target for the next generation of ultra high-frequency gravitational wave detectors.
%%%%%%%%%%%%%%%%%%%%%%%%%%%%%%%%%%%%%%%%%%
\begin{acknowledgments}
%%%%%%%%%%%%%%%%%%%%%%%%%%%%%%%%%%%%%%%%%%
\noindent
The work of S.V. was supported by the Kavli Institute for Cosmological Physics at the University of Chicago. MG was supported by the DGAPA-PAPIIT grant IA100525 at UNAM, and a Cátedra Marcos Moshinsky. A.G.V was supported by a SECIHTI (formerly CONAHCYT) graduate fellowship (CVU 2056403).
\end{acknowledgments}

\addcontentsline{toc}{section}{References}
\bibliographystyle{utphys}
\bibliography{references}

\end{document}